\documentclass[twocolumn,prb,showpacs,amsmath,amstex,amssymb,mathfonts,superscriptaddress]{revtex4-1}

\usepackage{amsthm,color,amsfonts,graphicx,verbatim}

\usepackage[dvipsnames]{xcolor}

\usepackage{amsmath}
\usepackage{amssymb}
\usepackage{amsthm}
\usepackage{amsfonts}
\usepackage{listings}
\usepackage{enumerate}
\usepackage{latexsym}
\usepackage{natbib}
\usepackage{epstopdf}
\usepackage{epsfig}
\usepackage{import}
\usepackage{psfrag}
\usepackage{bm}
\usepackage{graphicx}
\usepackage{dsfont}

\usepackage{hyperref}
\hypersetup{
    bookmarks=true,         
    unicode=false,          
    pdftoolbar=true,        
    pdfmenubar=true,        
    pdffitwindow=false,     
    pdfstartview={FitH},    
    pdftitle={Many-body localization in periodically driven systems},    
    pdfauthor={},     
    pdfsubject={},   
    pdfcreator={},   
    pdfproducer={}, 
    pdfkeywords={keyword1} {key2} {key3}, 
    pdfnewwindow=true,      
    colorlinks=true,       
    linkcolor=black,          
    citecolor=blue,        
    filecolor=magenta,      
    urlcolor=blue           
}

\theoremstyle{plain}

  \newtheorem{lemma}{Lemma}

\newcommand{\be}{\begin{equation}}
\newcommand{\ee}{\end{equation}}
\newcommand{\bea}{\begin{eqnarray}}
\newcommand{\eea}{\end{eqnarray}}
\newcommand{\HH}{{\cal H}}

\newcommand{\effcoup}{{\cal G}}

\newcommand{\e}{\mathrm{e}} 
 
\renewcommand{\i}{\mathrm{i}} 
\renewcommand{\d}{\mathrm{d}} 
\renewcommand{\phi}{\varphi}
\renewcommand{\epsilon}{\varepsilon}
\newcommand{\str}{ |}
\newcommand{\norm}{ ||}

\newcommand{\ipr}{\mathrm{IPR}}
\begin{document}
\title{Stability and instability towards delocalization in MBL systems}

\author{Wojciech De Roeck}
\affiliation{Instituut voor Theoretische Fysica, KU Leuven, Belgium}
\author{Fran\c{c}ois Huveneers}
\affiliation{CEREMADE, Universit\'e Paris-Dauphine, France}


%
\date{\today}
\begin{abstract}

We propose a  theory that describes quantitatively the (in)stability of fully MBL systems due to ergodic, i.e.\ delocalized, grains, that can be for example due to disorder fluctuations.   The theory is based on the ETH hypothesis and elementary notions of perturbation theory. The main idea is that we assume as much chaoticity as is consistent with conservation laws.  The theory describes correctly -even without relying on the theory of local integrals of motion (LIOM)- the MBL phase in 1 dimension at strong disorder.  It yields an explicit and quantitative picture of the spatial boundary between localized and ergodic systems. We provide numerical evidence for this picture. 

When the theory is taken to its extreme logical consequences, it predicts that the MBL phase is destabilised in the long time limit whenever 1) interactions decay slower than exponentially in $d=1$ and 2) always in $d>1$. Finer numerics is required to assess these predictions.

\end{abstract}

\maketitle

\section{Introduction} \label{sec: introduction}

The contrast between localized and thermalizing (ergodic) many-body systems is an exciting theme in condensed matter, touching upon foundations of thermodynamics.   The topic goes back to the seminal work \cite{anderson1958absence} but there has been revival due to some theoretical advances \cite{basko2006metal,gornyi2005interacting,oganesyan2007localization,pal2010many}, powerful numerics \cite{khemani2016obtaining,kjall2014many,luitz2015many} and the exciting experimental possibilities with cold atoms \cite{schreiber2015observation,smith2015many}. 
Most authors define many-body localization (MBL) as a property of the eigenfunctions at finite energy density: there is a full set of local conserved quantities: LIOM's 'Local Integrals of Motion' \cite{serbyn2013local,imbrie2016many,huse2014phenomenology}. We start from this definition even if it is surely not literally applicable to all systems that have been labelled as MBL (e.g.\ systems with a mobility edge, including in particular systems with no quenched disorder) and definitely not to all systems that exhibit MBL-like dynamical features\footnote{Although, in our opinion \cite{de2016absence}, systems with a mobility edge are also 'merely' MBL-like or glassy.}, e.g.\ classical disordered oscillator chains\cite{basko2011weak,huveneers2013drastic}.

\begin{figure}[h!]
\begin{center}
\includegraphics[width=8.5cm,height=6cm]{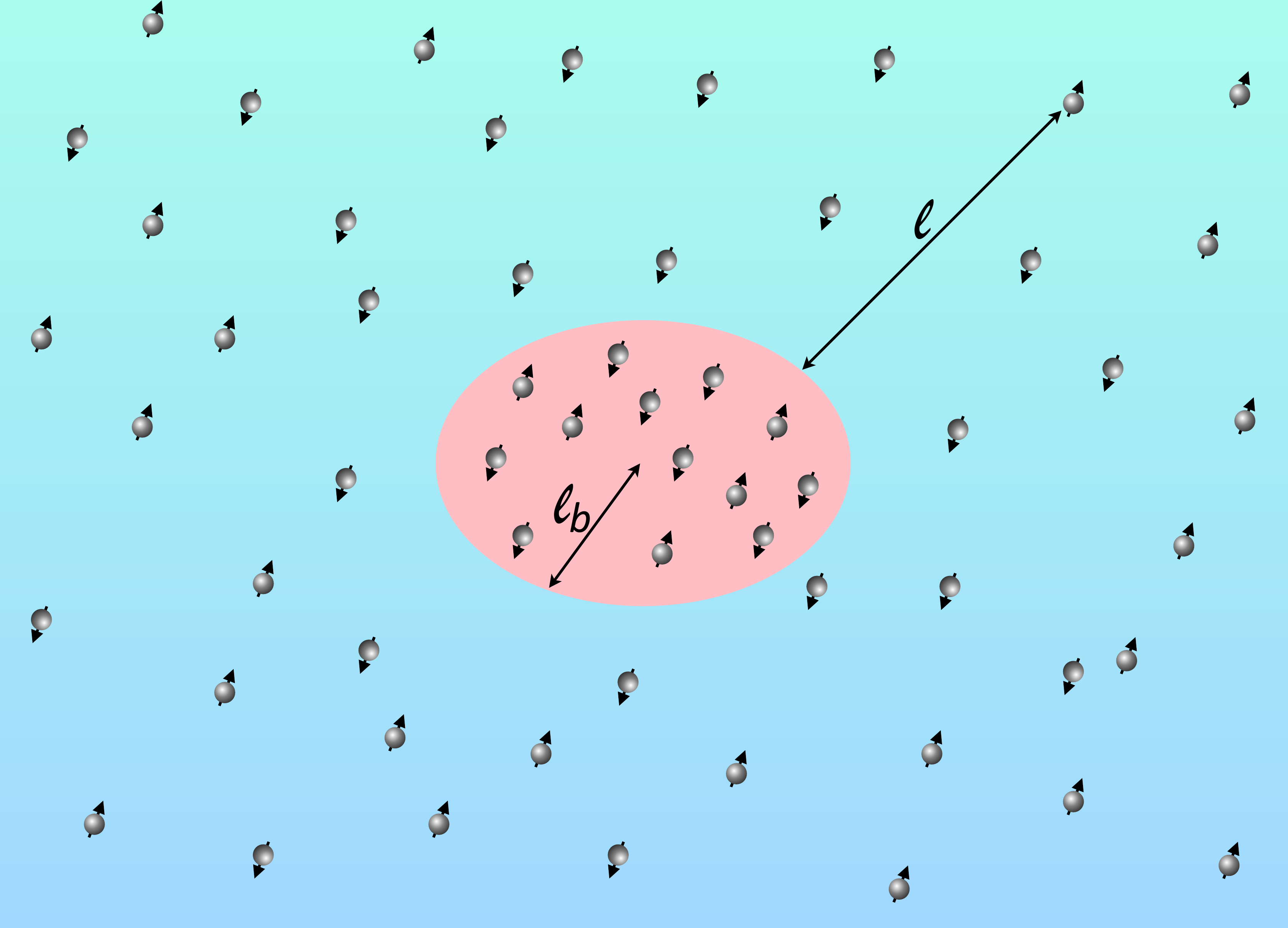}
\end{center}
\caption{The spins in the red region are closer to each other than in the surrounding blue-green region, hence they interact stronger. This stronger interaction  leads to an 'ergodic grain' in the red region. What is the effect of this grain on its localized environment? 
\label{fig: setup}}
\end{figure}

 This means in practice that our discussion is about strongly disordered lattice quantum systems.  We ask whether and how the localization in these strongly disordered systems is stable when we couple them to zero-dimensional, ergodic grains that are large but finite, in particular much smaller than the localized system. Such a setup occurs naturally in disordered systems where the ergodic grains arise from regions with anomalously weak disorder, see Figure \ref{fig: setup}.

The Hamiltonian of an ergodic system (or simply: a bath) is essentially described by random matrix theory (RMT) through the eigenstate thermalization hypothesis (ETH) \cite{deutsch1991quantum,srednicki1994chaos}.

Real space renormalization theories for MBL and ergodic inclusions  have been formulated \cite{potter2015universal,vosk2015theory,zhang2016many} but they are rather mesoscopic then microscopic and not formulated in the language of RMT.  This is at first sight logical since RMT is intrinsically connected with chaos and ergodicity. 
 Nevertheless, we point out that a simple RMT theory is able to describe the breakdown of  thermalization towards MBL in 1d.  
 In this point of view, localization emerges as an \emph{instability of RMT}. 
 
The motivation for our theory grew out of the bizarre observation that is by now well-known to many people in the field of MBL: the 'avalanche' or 'Ice-9' \cite{vonnegut1963cradle} scenario.
If a bath is coupled to a spin weakly, but strong enough to thermalize that spin, then the combined spin$+$bath system becomes ergodic and one is inclined to describe it simply by a random matrix. Doing so, the composed system looks like the original bath, except that its Hilbert space dimension is doubled and its level spacing is (roughly) halved. Therefore, it is now an even more powerful bath. Iterating this argument blindly one is led to conclude that any localization must be destroyed by a large enough ergodic grain that can serve as the original bath!  Obviously, we do not believe this scenario to be correct as it would rule out localization in strongly disordered $1d$ spin systems, where it has been almost proven \cite{imbrie2016many}. The RMT used in the present paper is subtle enough to avoid this fallacy by tracking the evolution of the structure factors (spectral functions) of the added spins. The narrowing of these structure factor eventually brings the avalanche to a halt in $1d$ systems. {More generally, our theory is able to describe correctly the stablity of strong localization to ergodic grains in all cases where we know the answer: $d=1$ interacting systems, and non-interacting systems.}
 
Yet, the most striking prediction of the RMT theory developed in the present paper is that the avalanche scenario has a core of truth: any degrees of freedom that are added to a bath and that do get effectively thermalized by the bath, also fully contribute to that bath. By this we essentially mean that it is the level spacing of the joint system, which is hence smaller than that of the bath alone, that determines whether other weakly coupled degrees of freedom get thermalized or not. This leads to a simple, universal theory of thermalization, where the bath is characterized by a single dimensionless parameter $\cal G$ that is essentially defined as $\textit{matrix element}/\textit{level spacing}$.   This should be contrasted with the question about timescales of thermalization or transport\cite{abanin2015asymptotic,parameswaran2016spin}, which is more complicated.

As a consequence, our RMT theory describes explicitly the spatial interface between an ergodic and a MBL system. In our theory, the cross-over region is fully thermal, but the structure (spectral) factors of operators supported in that region become very narrow as their support approaches the MBL system. 
 
 An intruiging prediction of our theory is that a localized system with interactions that decay slower than exponentially with distance, is not stable with respect to large ergodic grains, though the thermalization time for such a system can still be enormously large, thus making it localized for all practical purposes  
Counterintuitive as it might be, the idea that a few degrees of freedom
can delocalize a much larger number of localized degrees of freedom,
has by now been put forward already by several authors 
 \cite{aizenman2011resonant,chandran2016many,agarwal2015anomalous}.
Numerics for larger systems is however needed to nail down this instability.  
Also higher-dimensional localized systems would be unstable within our theory.

\subsection*{Outline}

{
In Section \ref{sec: bath spin problem}, we develop our random matrix theory for coupling spins to a bath. Then, in Section \ref{sec: stability against ergodic grains}, this theory is applied iteratively to the ergodic grains coupled to localized material.  Section \ref{sec: structure of the interface region} focuses on the interface region between ergodic grain and localized material.  We gather our numerical results and checks in Section \ref{sec: numerical tests}.
The appendices contain technical issues that can be skipped in a first reading.
}

\section{Bath+Spin Problem} \label{sec: bath spin problem}

Our method is to consider an ergodic system, that we call 'bath', coupled locally to small systems, $1/2$-spins for concreteness.  As such, our setup is very familiar from open quantum systems, see e.g.\ 
\cite{nandkishore2014spectral,johri2015many,levi2015survives,fischer2016dynamics} for specifics of baths coupled to localized systems.

 The point is however that we ask for detailed information on the eigenstates of the joint system, which will then be used to regard the joint system as a bath itself.  
For the sake of simplicity, we assume that energy is the only locally conserved quantity. We write the Hamiltonian as
$$
H={H_S}+{{{H_B}}}+  {{H_{SB}}},
$$ 
acting on $\HH_B \otimes \HH_S $ (`bath' $\otimes$ `spin'), and, as conventional, we slightly abuse notation by writing ${H_S}=1\otimes {{H_S}}$, ${{{H_B}}}= {{{H_B}}} \otimes 1$, etc.
To fix thoughts, we choose 
$$
{{H_S}}={{h}}\sigma_{S}^z, \qquad  {{H_{SB}}}= gV \otimes \sigma_{S}^x
$$
with $V$ a local operator acting in the bath (e.g.\ a $\sigma^x$ operator on the site adjacent to the spin $S$), a coupling strength $g$ and field ${{h}}$. We denote the eigenbasis of $H_S$ by $\str s\rangle$.  Additional interaction terms that do not flip the spin  re not essential at this stage.  In $d=1$ they are of course necessary to avoid integrability, but our scheme implicitly assumes throughout that the system is generic so this is not an issue, see also Section \ref{sec: general interactions}.

\subsection{Properties of the bath: ETH} \label{sec: properties of the bath}
Let us denote in general by  $\str b\rangle, E(b)$ the eigenvectors and energies of the bath Hamiltonian ${{{H_B}}}$. The important properties of the bath concern the matrix elements 
$V_{b,b'} =\langle b \str V \str b'\rangle$ in the $b$-basis. 
We assume the ETH  in the following form 
\be \label{eq: eth off}
V_{b,b'} = \frac{1}{\sqrt{{\rho}}} \sqrt{v(\omega)} \eta_{b,b'},  \qquad b\neq b' 
\ee
Here $\eta_{b,b'}$ are random variables of mean zero and unit variance, ${\rho}$ is the density of states at maximum entropy (see later), and $v(\omega)$ is a positive function of $\omega=E(b)-E(b')$ with units of $1/\text{energy}$, smooth on the scale of the level spacing.  
The dependence on the energy $E= (E(b)+E(b'))/2$ is less relevant since we naturally assume that $v$ depends smoothly on the \emph{energy density}  $\epsilon= E/N$, hence we omit it from the notation. 
The ETH also provides information on the diagonal elements, namely  
\be \label{eq: eth on}  V_{b,b}=\langle b \str V \str b \rangle= \langle V\rangle_{\epsilon}, \qquad   \langle b \str VV^* \str b \rangle = \langle VV^*\rangle_{\epsilon}, 
 \ee
where $\langle \cdot \rangle_{\epsilon}$ indicates the thermal ensemble at energy density $\epsilon=E(b)/N$. 
By subtracting constants, we can simplify and set $\langle V\rangle_{\epsilon}=0$ at the relevant energy density.  
By \eqref{eq: eth on},we have the following relation between the introduced quantities:
$$
\sum_{b'} \str V_{b,b'}\str^2 =   \langle VV^*\rangle_{\epsilon}
$$

If, as we primarily have in mind,  $V$ is of the simple form $V= \sigma^x_{B}$, then $ \langle VV^*\rangle_{\epsilon}=1$ if $\epsilon$ is at maximum entropy, as we will anyhow assume below.
  Using that $v$ varies on a scale much larger than the level spacing and that the density of  states is constant (see below), we can also write
\be \label{eq: sum rule}
\langle VV^*\rangle_{\epsilon}= \int \d \omega v(\omega), 
\ee
and likewise, we can identify $v(\omega)$ with the dynamic structure factor at zero momentum: 
$$
 \langle V(t)V^*\rangle_{\epsilon}
 =\int \d\omega \, v(\omega)  \e^{\i \omega t} $$
{for $t$ small enough so that single levels are not resolved, i.e.\ $\rho t\ll 1$, since we have assumed $v$ to be smooth.}
Treating $\rho=\rho(E)$ as a constant is in general not a valid approximation as $\rho(E) \propto \e^{s(\epsilon) N}$ where  the entropy density $s(\epsilon)$  varies smoothly with $\epsilon$. Hence $$\log(\rho(E+\Delta E)/\rho(E)) = s'(\epsilon)\Delta E+s''(\epsilon)(\Delta E)^2 /N, $$ 
the primes referrring to derivatives w.r.t.\ $\epsilon$.   From this we see that 
at maximum entropy, i.e.\ $ s'(\epsilon)=0$, we can yet treat $\rho(E)$ as a constant.    Generalizing our arguments to other values of the energy density is straightforward, and it brings no new insight, therefore we stick to the above setting. 

For baths with a local structure, the function $v$ decays rapidly at large $\omega$ \cite{mukerjee2006statistical}; $v(\omega) \leq C\e^{-\str\omega\str/\xi}$ can be proven rigorously \cite{abanin2015exponentially} with $\xi$ of the order of the total energy per site.  The behaviour at small $\omega \leq \xi$ reflects the long-time transport properties, i.e.\ $v(\omega) \sim \omega^{d/2-1}$ for a diffusive system and $v(\omega)=\text{const.}$ below the thouless energy, see \cite{d2015quantum, rigol2008thermalization,khatami2013fluctuation} for details. For our purposes these features are not important and for the moment, it suffices to think of $v$ as a simple bump function with halfwidth $\xi$.

\subsection{Basic RMT Hypothesis}\label{sec: basic hypothesis}

Our basic hypothesis describes the eigenstates of $H$ based on three  basic principles: 1) Hybridization condition (matrix element $\gg$ level spacing)  2) Energy conservation and 3) RMT is assumed whenever compatible with $1),2)$. \\
 Let us make this more precise. The energy difference between the 2 spin states $s=\pm 1$ is $2h$, so the relevant matrix element of the perturbation is of size $g \sqrt{v(2h)/\rho}$, by \eqref{eq: eth off}, whereas the level spacing between states with opposite spin $s$ is $1/{\rho}$. Hence the condition for hybridization is 
\be \label{eq: hybridization}
{\effcoup} \equiv g \sqrt{ v(2h) { \rho}}   \gg 1.
\ee
If the hybridization condition is not satisfied, then the spin is not thermalized by the bath and the coupled eigenstates are determined by perturbation theory. If the hybridization condition \emph{is} satisfied then we propose a form for the new eigenstates, namely
$$
\psi =  \frac{1}{\sqrt{ {2\rho}}}  \sum_{b,s } \sqrt{k(\omega)} \,  \eta(s,b) \,    \str s, b \rangle, \qquad  \omega = E(s,b)-E
$$
As before, the $\eta(s,b)$ are i.i.d.\ random variables with zero mean and unit variance. $E(s,b)=hs+E(b)$ is the energy of the eigenstate $\str s,b \rangle=\str s \rangle \str b \rangle$ with respect ot the uncpupled Hamiltonian $H_S+H_B$.
 The \emph{hybridization function} $k\geq 0$ is assumed to be smooth on the scale of the level spacing $1/\rho$. It satisfies the normalization $\int \d\omega k(\omega)=1$, ensuring that $\norm \psi\norm=1$, and it has dimension $\text{Energy}^{-1}$. The factor $2\rho$ is the density of states of the $B+S$ system (recall $\rho$ is the density of the bath).
This ansatz expxresses that the new eigenstates are random superpositions of uncoupled eigenstates, up to energy conservation which is expressed by the fact that the function $k$ is cut-off at large $\str\omega\str \geq w$, where we have defined a \emph{halfwidth} $w$.  It is natural to assume that for $\str\omega\str \leq w$, the function $k$ is essentially flat.  In the next section, we refine the information on the function $k$, but for most of our applications such detailed information is not necessary: all that really matters is that $w$ does not depend on the total Hilbert space dimension (or, equivalently, not on $\rho$). 
\subsection{The hybridization function $k$}\label{sec: hybridization function}

First, let us deduce the value of the halfwidth $w$ introduced above, i.e.\ the energy range up to which old uncoupled eigenstates become completely mixed.  This is given by the Fermi Golden rule (FGR)  $w = g^{2} v(2h)$ provided that the rate $w$ is smaller than the correlation time in the bath. In practice, this means that $w$ needs to be smaller than the characteristic $\omega_0$ over which $v$ changes appreciably around $\omega= 2h$.  In the present paper, the FGR will always be applicable, at least when taking into account a caveat that we postpone to Section \ref{sec: correction}. 
{Note that as long as $\xi$ is not smaller than $2h$, the approximation $v(2h)\approx 1/\xi$ is reasonable and  we have hence simply $w \approx g^{2}/\xi$.}
 For the most elaborate application of our theory, we need also the form of $k$ for  intermediate $w\ll\str\omega\str\ll \xi$ (for $\str\omega\str\geq \xi$, it is easy to argue that $k$ inherits the exponential decay of $v$). Simple reasoning based on perturbation theory (see also Section \ref{sec: correction}) yields that $k(\omega) \sim \omega^{-2}$ and it is therefore natural to propose for $k$  a Lorentz shape
$$
k(\omega)=    \frac{1}{\pi w (1+(\omega/w)^2)}
$$
This shape neglects the exponential falloff for $\str\omega\str\gg \xi$ which is not essential for our results (it can of course easily be incorporated). Numerical tests, see Appendix \ref{app: numerical evidence for hybridization width}, confirm the the FGR expression for the width $w$ and the asymptotic $k(\omega) \sim \omega^{-2}$. Figure \ref{fig: Hybridization Hamiltonian bath} shows the hybridization curve for a real bath, we see that that curve has some structure for $\str\omega\str \leq w $, so the Lorentzian form is merely an idealization.


\subsection{New structure factor}\label{sec: new structure factor}

Let us determine the matrix elements of a bath operator $V$ upon adding the spin when the new eigenfunctions are as described in the previous section via a hybridization function $k$. This is a straightforward calculation starting from
\be \label{eq: v bath abstract}
V_{\psi,\psi'} = \sum_{s,b,b'}  \langle \psi\str b,s \rangle  V_{b,b'}   \langle b',s \str \psi' \rangle
\ee
where we used that $V$ acts on the bath only and $\psi,\psi'$ are eigenfunctions (at energies $E,E'$) of the coupled system. 
We use that $V$ had a vanishing diagonal in the $b$-basis and, at least at maximal entropy, this remains so in the new basis\footnote{Up to corrections vanishing in the thermodynamic limit}.
For the off-diagonal, we replace $V_{b,b'}, \langle \psi\str b,s \rangle, \langle b',s \str \psi' \rangle$ by their explicit (random) expressions. 
Pretending that all random variables are independent, we get that the variance of $V_{\psi,\psi'}$ is given by 
$$
\langle \str V_{\psi,\psi'}\str^{2}\rangle  =2 \frac{1}{4\rho} \int \d \omega_1 \d\omega_2  \,  k(\omega_1)k(\omega_2)   v(\omega-\omega_1-\omega_2)
$$
where $\omega=E-E'$ and the factor $2$ in front originates from the sum over the spin variable $s$.  Making again an independence assumption, we conclude that the new system satisfies again ETH in the sense of \eqref{eq: eth off} with density $\rho'=2\rho$ and structure factor 
\be \label{eq: updated v in bath}
v'(\omega) =  (k \star v \star k)(\omega)
\ee
where $(a\star b)(\omega)\equiv \int \d \omega_1 a(\omega_1) b(\omega-\omega_1)$ is convolution of functions. 

The upshot is that the function $v'$
is given by $v$ smoothened twice with halfwidth $w$. In particular, if $g^2/\xi \ll \xi$, which is the case we have in mind, then the two functions $v'$ and $v$ are nearly the same.  Of course, as the density of states $\rho$ is doubled, this means that the typical matrix element of $V$ did become smaller by a factor $\sqrt{2}$. 
On the one hand, this is a very intuitive conclusion: the new spin has made the bath more powerful by increasing its effective dimension. 
On the other hand, the conclusion that the structure factor of a bath operator is unaffected by the coupled spin that is potentially much slower than the bath correlation time, does not hold up to scrutiny\cite{altmanbergprivate}.  Indeed, the correlation function $\langle V(t)V\rangle_{\epsilon}$ should acquire long lived oscillations with frequency $\omega=\pm 2h$ due to the slow spin. Below, in Section 'Backreaction Correction', we explain how this feature emerges within our formalism upon refining the RMT assumption. Since this does not affect our results, we however ignore this in the rest of the paper.

Finally, let us now also determine the structure factor of an operator on the coupled spin, $V=\sigma^x_S$, i.e.\ $\langle b,s \str V \str b',s' \rangle=\delta_{b,b'}\delta_{s,-s'}$. Instead of \eqref{eq: v bath abstract}, we start now from
\be \label{eq: v spin abstract}
V_{\psi,\psi'} = \sum_{s,b}  \langle \psi\str b,s \rangle  \langle b,-s \str \psi' \rangle
\ee
Using analogous steps as the derivation of \eqref{eq: updated v in bath}, we find for the new structure factor
\be \label{eq: v spin}
v'(\omega) = \frac{1}{2} (k \star k)(\omega+2h) + \frac{1}{2} (k \star k)(\omega-2h),
\ee
This represents two bumps centered at $\pm 2h$ of halfwidth $2w$.  Note that $\int v'=1$ (since $\int k=1$), which is consistent with $\int v'=\langle VV^* \rangle_{\epsilon}$, see \eqref{eq: sum rule}.
{The expression \eqref{eq: v spin} can simply be obtained from \eqref{eq: updated v in bath} by putting $v(\omega)=\frac{1}{2} \delta(\omega+2h) + \frac{1}{2}\delta(\omega-2h)$, which can indeed be considered as the structure factor of $\sigma^x_S$ with respect to the uncoupled Hamiltonian $H_S+H_B$.}

\subsection{Backreaction Correction}\label{sec: correction}

As already remarked above, our theory misses a backreaction effect that creates spikes in the bath structure factor. We investigate now a refinement of the theory that does allow to recover those spikes.

We write in general $H_0=H_S+{{H_{B}}}$ and the corresponding energies as $E(b,s)$.
  We choose a target energy $E$ and we let $P$ be the spectral projection associated to $H_0$ of the interval $[E-w,E+w]$ with some halfwidth $w$ that we will assume to be the same as the halfwidth introduced above, even though this is not necessary for the upcoming lemma.  Also, we write $\bar P=1-P$.

Our main tool is the following simple principle, whose proof follows directly from the eigenvalue equation 
\begin{lemma}[Schur complement formula] \label{lem:schur}
 Let $\psi$ be an eigenvector of $H$ with eigenvalue $E$ and $P\psi\neq 0, \bar P\psi\neq 0$, then $\bar P (E-H)  \bar P$ is invertible on $\bar P {{H_{SB}}} P \psi$ and
\be \label{eq: schur one}
 \bar P \psi=   \frac{1}{\bar P(E- H) \bar P} \bar P {{H_{SB}}} P \psi 
\ee
\end{lemma}

This formula expresses a useful structural relation between the spectral regions close to and far from the energy $E$. 

An obvious way to take this relation into account is to retain the random matrix form of the eigenstates $\psi$ inside the interval $[E-w,E+w]$, i.e.\ for $P\psi$, but to relate $\bar P \psi$, i.e.\ outside the interval $[E-w,E+w]$, to $P\psi$ by the above Lemma \ref{lem:schur}. 
This means that we no longer assume that all $\eta(b,s)$ in Section \ref{sec: basic hypothesis} are independent, but just those inside the interval $[E-w,E+w]$. Hence, we write $\psi=P\psi+\bar P \psi$ and we postulate
\be \label{eq: refined proposal}
P\psi := \frac{1}{\sqrt{2\rho}}\sum_{b,s} \sqrt{k_0(\omega)} \eta(b,s) \str b,s\rangle 
\ee
where again $\omega=E-E(b,s)$, $\eta(b,s)$ are independent random variables with mean zero and unit variance, and we have introduced a cut-off hybridization function $k_0(\omega):= \chi(\str\omega\str \leq w) k(\omega)$.  The remainder $\bar P \psi$ is then obtained by \eqref{eq: schur one}. This model for $\psi$ refines the proposal in Section \ref{sec: basic hypothesis}.
Within this model, let us now calculate $v'(\omega)$, the updated structure factor of $\sigma^x_B$, i.e.\ we refine the result of Section \ref{sec: new structure factor}. We start from the decomposition 
\begin{align}
\langle \psi \str \sigma^x_B \str \psi' \rangle  =&  \langle P\psi \str \sigma^x_B \str  P' \psi' \rangle+ \langle P\psi \str \sigma^x_B \str 
 \bar P'\psi' \rangle \nonumber \\
 &+\langle \bar P\psi \str \sigma^x_B \str 
  P'\psi' \rangle +\langle \bar P \psi \str \sigma^x_B \str \bar P'\psi' \rangle \label{eq:inandout}
\end{align}
Here the primed projectors $P',\bar P'=1-P'$ are associated to $\psi'$; i.e.\ they are centered on $E'$ instead of $E$. 
The new effect can be seen most easily on the second (or third) term.  
 Using Lemma \ref{lem:schur} and plugging ${{H_{SB}}}=g \sigma^x_S \sigma^x_B  $, we get
\begin{align}
\langle P\psi \str \sigma^x_B \str \bar P'\psi' \rangle &=\sum_{b,b',s,s'} \langle P\psi \str b,s \rangle  K(b,s \str b',s')   \langle b',s' \str  P'\psi' \rangle  
\end{align}
with the first and last factor given still by the RMT ansatz and 
$$
K(b,s \str b',s') = g \langle b,s\str  \sigma^x_B  \frac{1}{\bar P' (E'-H)\bar P'} \bar P' \sigma^x_S \sigma^x_B  \str b',s' \rangle  
$$
Whereas we know that the diagonals $\langle\psi\str\sigma_B^x\str\psi \rangle$ vanish (see Section \ref{sec: new structure factor}), there is in general no reason why $K(b,s \str b',s')$ should vanish for $b=b'$. These partially diagonal terms yield contributions to the structure factor $v'$ that are not directly related to $v$ but are instead peaks of halfwidth $2w$ around $\omega=0$ and $\omega=\pm 2h$. To be specific, up to lowest non-vanishing order in $g$, we find, see Appendix \ref{app: backreaction}, the following contribution to $v'$:
\be
\mathcal{W} \,  k_0 \star  \delta(\cdot \pm 2{{h}}) \star k_0,\qquad  \mathcal{W}\equiv (\tfrac{g}{\max{\xi,\str h\str}})^2  \label{eq: diagonal contribution v}
\ee
I.e.\ two peaks with weight $ \mathcal{W}$ and halfwidth $2w$, from the smoothing with $k_0$ (the fact that we get here $k_0$ instead of $k$ is most likely irrelevant). These peaks are located at the Bohr frequencies of the external spin $\pm 2{{h}}$, fully in line with the intuition from the time-domain. 

The contributions from $K(\cdot)$ with $b\neq b'$,
 together with the first term in \eqref{eq:inandout}, basically recreate the previously found  form of $v'(\omega)$, up to normalization (see below). In fact, the calculations coming here can also be used to justify the choice of the width, i.e.\ the FGR, see Appendix \ref{app: backreaction}. 

Putting all pieces together, we find within our refined model, and up to lowest nontrivial order in $g$, that the structure factor $v'$ is given by 
\be
(1-2\mathcal{W})\, k\star v\star k +  \mathcal{W}\, k_0 \star    \delta(\cdot \pm 2{{h}}) \star k_0
\label{eq: refined structure factor}
\ee
The factor $(1-2\mathcal{W})$ is due to the overall normalization $\int v'=1$. In the relevant case $w\ll \xi$, the new peaks are less smooth than the structure factor of the original bath, but then $\mathcal{W}\ll 1$. Hence, then 
the strength of the smooth part of the bath has been slightly depleted by the appearance of narrow peaks.  

Let us discuss the implications of the refinement presented above as we will apply the theory iteratively, in Section \ref{sec: stability against ergodic grains}. 
First, as we will now be dealing with a situation where the structure factor $v$ is a sum of a smooth part $v_{\text{sm}}$ and a more irregular part with narrow spikes $v_{\text{irr}}$, with $\int v_{\text{irr}} \ll \int v_{\text{sm}} $, we have to reconsider the reasoning in Section \ref{sec: hybridization function} on the hybridization width.  In such a case, it can happen that the Fermi Golden Rule is applicable for $v_{\text{sm}}$ but not for $v$ itself and then we have $w \approx g^{2} v_{\text{sm}}(2h)$ (a superficial justification of this is given in Appendix \ref{app: backreaction})

Secondly, since we will couple spins with rapidly decaying couplings $g_i$ to the same bath, we will get a depletion of the smooth structure factor by the factor $\prod_i (1-2{\cal W}_i)$ with $\mathcal{W}_i\equiv (\tfrac{g_i}{\max{(\xi,\str h_{i}\str})})^2$. Because of the rapid decrease of $g_i$, this factor will still be close to $1$ and so the overall depletion effect remains small.  

Hence, the conclusion is that the refinement proposed in this section, does not have any implications for us, and we will henceforth ignore it.

\subsection{General interactions}\label{sec: general interactions}

What if, in addition to the coupling term $g\sigma_B^x\otimes \sigma_S^x $, there is also a coupling term of the form, say,  $g'\sigma_B^z\otimes \sigma_S^z $?  Such terms will in general be present and in the one-dimensional case, they are even necessary to avoid a trivial integrability when building up a chain. It is easy to come up with a generalization of our basic RMT rules to handle such terms. Given a Hamiltonian $H=H_0+V$, not necessarily of system+bath form, with $H_0$ ergodic and the perturbation $V$ local, we can postulate that the eigenstates of $H$ are formed as random superpositions of the eigenstates of $H_0$, with again a hybridization function $k$ whose width is determined via the FGR. 
This philosophy can be applied in more than one way, but the following seems to us in general the most accurate: In a first step, we add the  $g'\sigma_B^z\otimes \sigma_S^z $ term to $H_S+H_B$ and then, in a second step, the $g\sigma_B^x\otimes \sigma_S^x $ term. 
For the first step, we can fix the value of the external spin $s$, since the coupling term commutes with $\sigma_S^z$. Hence we get in fact two different problems, depending on $s$, with perturbation term $V_s= g's\sigma_B^z$. We treat these two problems by the general philosophy explained above (note that in general there might be a diagonal term that results in a simple $s$-dependent shift). The second step proceeds as before, but for this step the unperturbed eigenstates are no longer products.  The calculations are complicated and the whole generalization adds little to our theory, except for eliminating some traces of integrability. Hence, we will completely ignore these general interaction terms in the present paper.

\subsection{Limitations of our theory}\label{sec: limitations}
Our theory makes some uncontrolled assumptions. 
In particular, apart from the refinement introduced in Section  \ref{sec: correction}\footnote{
This correction is not sufficient to solve the problems addressed below, and will be ignored here.}, 
we assume that the matrix elements of a bath operator $V$ as well as the parameters $\eta(b,s)$ featuring in the hybridization function are all mutually independent. 
This assumption is definitely the main source of non-rigor.\\
\noindent\emph{Locality} In effect, our theory provides an expression for the eigenstates of the coupled system as a random superposition of the states of the uncoupled system, taking into account conservation laws. 
Upon iterating, these putative eigenstates are random superpositions of products over sites.
This picture can never capture locality.
The same problem occurs if one would use the Berry conjecture \cite{berry1977regular} (eigenstates are random superpositions of plane waves with appropriate momenta) as a cartoon for ETH in many-body systems. The rapid decay of the structure factor $v(\omega)$ as $\omega\to\infty$ is not captured by this cartoon and it needs to be imposed explicitly.  Also our theory sometimes misses the decay of structure factors: 
If we build up an ergodic chain of lenght $\ell$ by coupling spins with $g\sim h\sim \xi$, 
then our theory predicts that each coupled spin broadens by a similar amount the structure factors and eventually some local operators have a width of order ${\ell}$, analogously to a featureless random matrix with the same bandwidth.
 More generally, this issue leads to a clear error if the total broadening is comparable or larger than the original width $\xi$ of bath operators, which occurs -roughly speaking- if
 $\sum_i g^2_i \geqslant \xi^2 $ with $g_i$ the coupling strengths of added spins.  
 For our purposes, we can largely circumvent this issue. 
Firstly, resorting to the LIOM representation for the localized part of the system, we can encode most of the locality through the exponential decay in space of the coupling of LIOMs to the bath and so indeed the $g^2_i$ are rapidly decaying
(the problem of encoding the locality inside the bath and its close vicinity remains unresolved in that way, but this leads to rather minor corrections). 
Secondly, in $d=1$, this problem is so mild that we will even manage to proceed without using the LIOM theory, see Section \ref{sec: stability of mbl without}.\\ 
\noindent\emph{Proximity effects} These are effects 
\cite{hyatt2016many,nandkishore2015many} whereby a localized system localizes the bath by inducing effective disorder terms via the coupling.  These effects arise when the coupling to the localized system dominates the ergodic bath Hamiltonian.
If we keep coupling spins with coupling strenghts $g_i$ directly to a finite bath, then proximity effects will occur when $(\sum_i g^2_i)^{1/2}$ becomes comparable to the total bath energy but our theory is unable to detect this.    This scenario is however not realized in the systems studied in this paper.

 \section{Stability against ergodic grains}\label{sec: stability against ergodic grains}
 
\subsection{Stability of one-dimensional localization}\label{sec: stability one dimension}

\begin{figure}[h!]
\begin{center}
\includegraphics[width=7cm,height=2cm]{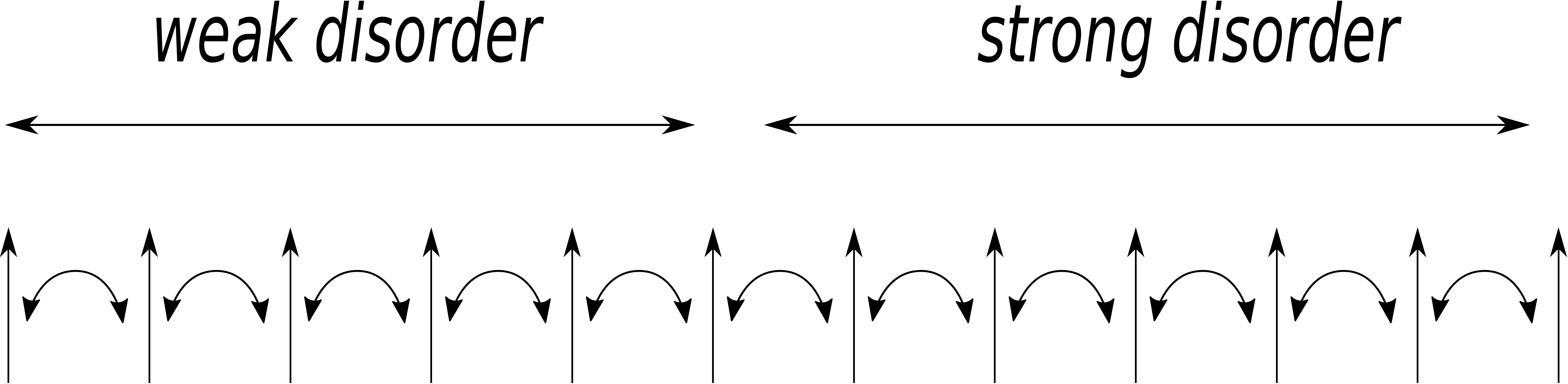}\\ 
\includegraphics[width=7cm,height=1.7cm]{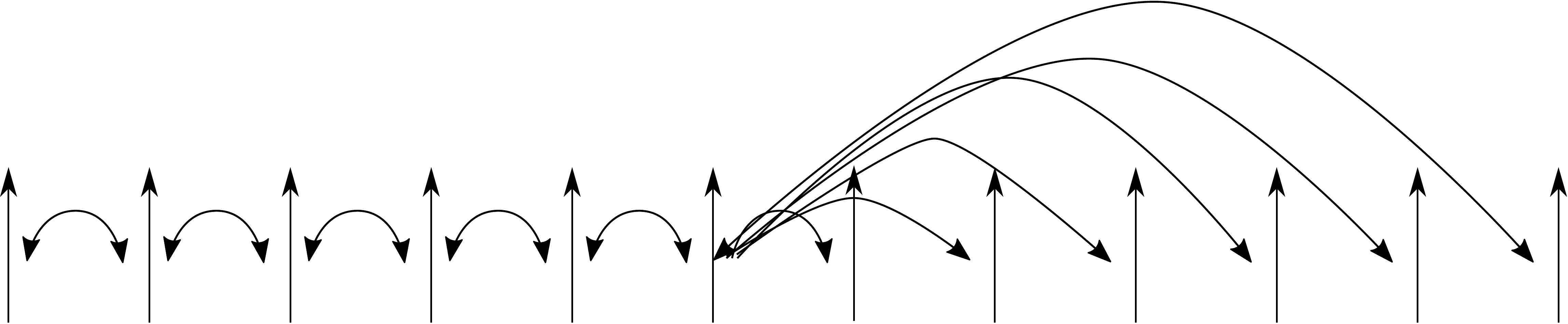}\\ 
\includegraphics[width=7cm,height=1.7cm]{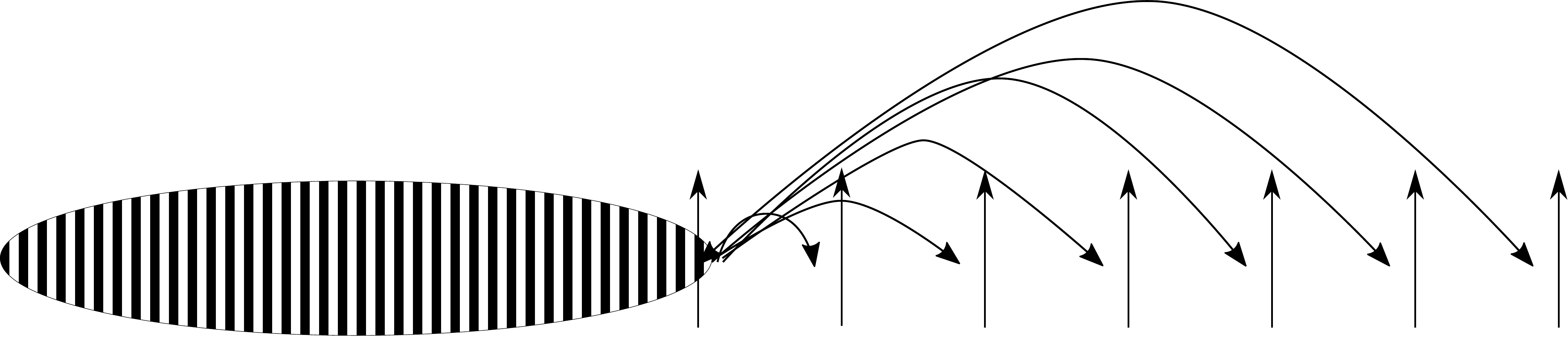}  
\end{center}
\caption{
\label{fig: modelling}
\emph{Top}: A chain of weakly disordered spins -the ergodic grain- is coupled to a chain of strongly disordered spins -the MBL system.
\emph{Middle}: The system is modelled by weakly disordered spins coupled to LIOMs (no coupling between the LIOMs any more). Note that all LIOM's are coupled only to the rightmost spin.
\emph{Bottom}: The weakly disordered spins are sometimes modelled by a random matrix
}
\end{figure}

We apply our theory to a bath of $\ell_b$ spins attached to  a strongly localized spin chain of length ${\ell}$ at its right. 
However, instead of considering ${\ell}$ weakly coupled disordered spins, we invoke the LIOM theory and we consider ${\ell}$ uncoupled l-spins. The l-spin operators are denoted by $\tau^{x,y,z}_i$ instead of $\sigma^{x,y,z}_i$, and the $\tau^z_i$ commute with the Hamiltonian of the ${\ell}$-stretch.   This Hamiltonian does hence not contain any more terms that flip the $\tau^z$.  However, the -strictly local- coupling of the leftmost of the ${\ell}$ spins to the bath, gives rise to a nonlocal, though exponentially decaying, coupling of the l-spins to the bath, see Figure \ref{fig: modelling}.

This setup is captured by the Hamiltonian
\be \label{eq: ham bath plus loc}
H={{{H_B}}} + \sum_{i=1}^{{\ell}}  h_i \tau^z_{i}  +        \sum_{i=1}^{{\ell}}  g_i  \sigma^x_0 \otimes \tau^x_{i}, 
\ee
with decaying couplings 
$$
\qquad g_i= g_1\e^{- (i-1)/\zeta}   
$$
where $\zeta$ is a possible definition of the localization length (in units of the lattice spacing).
Note that all the LIOM's are coupled to the same bath operator $\sigma^x_0$ and that $g_1$ is the coupling strength of the leftmost physical spin to the bath. The following kind of terms were suppressed in our model Hamiltonian \eqref{eq: ham bath plus loc}:
1) nonlocal LIOM-energy terms like e.g.\ $\tau^z_i \tau^z_{i+1} $, and 
2) bath couplings affecting multiple LIOMs like $\sigma^x_0 \otimes \tau^x_1 \ldots\tau^x_{i}$.
These terms should be present generically but we omit them since they don't change  
 qualitatively the reasoning that follows.  We take $h_i \sim \xi$ with $\xi$ the halfwidth of the structure factor $v$ of the bath operator $\sigma_O^x$. This is realistic if the bath is made up from similar spins as the localized chain. Of course, the $h_i$ should be disordered but this is not important any longer in \eqref{eq: ham bath plus loc}: the strong disorder has already been used to derive this model Hamiltonian, and it is now  by $g_1 \ll \xi$ and $\zeta \ll 1$.

We proceed inductively, setting first $g_i=0$ for $i > 1$.  That puts us precisely in the case discussed at length in Section \ref{sec: bath spin problem}, with $V= \sigma_0^x$. The hybrdization condition \eqref{eq: hybridization} will be clearly satisfied if ${\ell_b}$ is large enough and we obtain eigenstates $\psi$ of the $B+S_1$ system. The next step is to view the $B+S_1$ system as a the bath and to couple it to $S_2$ through 
$V= \sigma_0^x$.  If  ${\ell}$ was large enough, the hybridization condition is agains satisfied and we can proceed. The only way that our scheme can stop is if at some point the hybridization condition is violated. 
To evaluate the hybrdization condition we need to determine at each step the new density of states $\rho'$ and structure factor $v'$ from those at the previous step $\rho, v$. Naturally, $\rho'=2 \rho$ and from  \eqref{eq: updated v in bath} in Section \ref{sec: new structure factor} we deduce that, roughly, $v' \approx v$: the structure factor stays roughly the same since the sum of widths of all the hybridization functions is small compared to the width of $v$:  $\sum_i g^2_i/\xi \ll \xi$.  This means that the ${\effcoup}$-parameter defined in \eqref{eq: hybridization} gets updated as
$$
{\effcoup}'=  {\effcoup} \e^{-a}, \qquad a \equiv 1/\zeta-\log(2)/2 \gg 0
$$
where the inequality $ a \gg 0$ follows from strong disorder $\zeta \ll 1$. 
It follows that ${\effcoup}_{{\ell}}=\e^{-a{\ell}} {\effcoup}_1$. Since ${\effcoup}\approx  (g_0/\xi) 2^{\ell_b/2} $, by ETH for the original bath, we conclude that the hybridization condition breaks down at $\ell=\ell_c$ with
\be {\ell_c} \approx \frac{\log{2}}{2/\zeta-\log{2}} \ell_b \label{eq: length crossover} 
\ee
and this length is  hence an estimate for the size of the crossover region.  In a more general estimate, the factor  $\log{2}$ should of course be replaced by an entropy density. 

Note that our estimate differs from the more simple guess whereby one considers only the thermalizing effect of the original bath with length $\ell_b$; such an estimate leads to a crossover region of size 
$$\ell_c\approx\zeta \frac{\log{2}}{2} \ell_b.$$   More importantly, our estimate suggests that LIOM's can not have an arbitrarily large localization length, because if $2/\zeta \leq\log{2}$, then the localization is not stable with respect to ergodic grains. 
One should take care to interpret this statement correctly. It does not contradict the fact that the localization length diverges at the transition from MBL to ergodicity. It simply means that close to the transition the increase in localization length is due to the proliferation of resonances, rather than to a change in the structure of resonance-free regions.

\subsection{Instability of MBL for subexponentially decaying interactions}\label{sec: instability subexponential}

The above analysis directly implies that a spin chain with subexponentially decaying interactions is not stable wit respect to ergodic grains. Indeed, if such a system were MBL, then the LIOM operators should presumably have a subexponential tail as well \footnote{The more precise statement should be that terms that affect only two LIOM's are subexponentially decaying in the distance. Terms that affect $n$ LIOMs decay in addition exponentially in $n$} and we can model their interaction with an ergodic grain by taking the $g_i$ in \eqref{eq: ham bath plus loc} to decay subexponentially.  In this case, however, the ${\effcoup}$-parameter flows to infinity if $\ell_b$ is large enough, i.e.\ if the grain is sufficiently large. 
A tentative step towards verifying this is described in Section \ref{sec: numerical tests}, where we consider a random matrix corresponding to a 6-spin bath coupled weakly to a chain of 8 LIOM spins with decay factor $\e^{-1/\zeta}=g_{i+1}/g_i=3/4$, for which $\ell_c=\infty$ according to the above estimates. Despite the LIOM chain being localized, we see that the resulting system indeed behaves rather accurately as an ergodic system with dimension $2^{6+8}$.

Finally note that if we would change the model, so as to have very slowly decaying $g_i$, then proximity effects, see Section \ref{sec: limitations}, can indeed localize the ergodic grain.  We do not discuss this as such models can probably not emerge as LIOM's of local localized Hamiltonian\cite{burin2015many,hauke2015many}.

\subsection{Instability of MBL for higher-dimensional systems}\label{sec: instability higher dimensions}

The higher-dimensional setup is in effect similar to the case of sub-exponentially decaying couplings. 
Consider a spherical ergodic grain of with radius $\ell_b$, surrounded by LIOM's coupled to it with strength decaying exponentially in distance $r$, $g_r \approx g_0 \e^{-r/\zeta}$. The nearby LIOM's will get thermalized and according to our theory above, any thermalized LIOM fully contributes to the bath.   
The number of  LIOM's with distance  $r \leq \ell$ is
$$
N_\ell= C_d (\ell_b+\ell)^d-\ell_b^d
$$
with $C_d$ the volume of a unit $d$-dim sphere.
When these LIOM's have been thermalized, the bath density of states has been increased by a factor $2^{N_{\ell}}$, which grows superexponentially in $\ell$ if $d>1$. This overwhelms hence the effect of exponentially decreasing couplings. 
Said a bit differently, if there were a crossover region extending up to a distance $\ell_c$ from the bath, and beyond that region MBL would persist, then there is a thermal volume (bath+crossover region)
$V_{th}(\ell_c) =C_d (\ell_{b} + \ell_c)^d$. The condition that spins outside the ergodic region are not hybridized is then
\begin{equation}\label{size thermal volume}
V_{th}(\ell_c) \log 2 - \frac{2}{\zeta} l_c \leq 0, 
\end{equation}
This equation has a finite solution for $\ell_c$ either when $\ell_b$ is small enough compared to $\xi$, or for $d=1$ as soon as\footnote{In the previous sections, the condition $\xi < 2/\log 2$ was derived instead. This is simply because there the localized material was only on one side of the ergodic grain, whereas now it is on two sides.} $\xi < 1/\log 2$.  

{Of course, this estimate is known already as an upper bound on the cross-over region; it is for example the main reason why the analysis of \citep{imbrie2016many}, in which the cross-over region is called 'collar' is restricted to $d=1$. The point in the present paper is however that within our theory, the volume $V_{th}$ is not an upper bound, but the actual volume of a fully ergodic region. }

As a note of caution, we remark that the $d>1$ setup is not entirely free of the problem discussed in Section \ref{sec: limitations}\emph{locality}. Already when coupling the first layer of spins around a spherical ergodic grain of diameter $\ell_b$, the structure factors in the grain grow like $\ell_b^{d/2}$ if we were to apply our theory literally.

Finally, we note that our conclusions echo the analysis in \cite{nandkishore2014spectral}, where a careful investigations of structure factors led to a division  of MBL systems in 'weak' and 'strong' MBL, where only $d=1$ systems with exponentially decaying interactions can be 'strong MBL'. Our analysis suggests however that 'weak MBL' systems are delocalized.

%
%
%
%

\subsection{Stability of MBL without LIOM's}\label{sec: stability of mbl without}

In this section, we rederive the stability of one-dimensional MBL that was established in Section \ref{sec: stability one dimension}. However, we do not use the powerful LIOM representation for the localized spins. Instead, we develop our reasoning here for physical spins instead of l-spins. The possibility of doing so demonstrates the versatility of the theory. 

Hence, we consider again the Hamiltonian \eqref{eq: ham bath plus loc}, but now written in terms of the physical $\sigma$-operators:
\be
H={{{H_B}}} + \sum_{i=1}^{{\ell}}  h_i \sigma^z_{i}  +    g   \sum_{i=1}^{{\ell}}  \sigma^x_{i-1} \otimes \sigma^x_{i}
\ee
with again $\sigma^x_{0}$ a bath operator, cf.\ \eqref{eq: ham bath plus loc}, and $h_i$ random fields, i.i.d.\ random variables uniformly drawn from $[-h,h]$ with $g/h \ll 1$, i.e.\ strong disorder.
Let us denote by $v_i, i=0,1,\ldots$ the structure factor of the operator $ \sigma_i$, these are relevant because they couple to spin $i+1$. For $v_0$ we take a bump function with halfwidth $\xi \sim h$ (as before), and the other $v_i$ are to be determined. 
We write $w_i$ for the width of the hybridization function by which the $i$'th spin is coupled to the preceding ones. 
So, assuming the FGR applies (see below) we have
\be \label{eq: width i}
w_i = g^2v_{i-1}(2h_i)
\ee
The structure factors $v_i, i\geq 1$ are given through \eqref{eq: updated v in bath} as
\be \label{eq: structure i}
v_i(\omega)= \frac{1}{2\pi w_i} \big(\frac{1}{1+(\frac{\omega + 2h_i}{2w_i})^2}+ \frac{1}{1+(\frac{\omega - 2h_i}{2w_i})^2}\big)
\ee
because the convolution of two Lorentz distributions with halfwidth $w_i$ is again a Lorentz distribution with halfwidth $2w_i$. 
The above two relations \eqref{eq: width i} and \eqref{eq: structure i} in determine inductively all structure functions. Let us neglect resonances by pretending that all neighboring fields are sufficiently different: 
$$\min{ \str h_i-h_{i-1}\str, \str h_i+h_{i-1}\str }\approx h/2, $$ 
Keeping then only the largest of two terms in \eqref{eq: structure i} and using that $h/w_i \gg 1$, we get
\be \label{eq: width i explicit}
w_i=g^2 v_{i-1}(2h_i) \approx \frac{g^2 w_{i-1}}{\pi h^2}
\ee

 So the upshot is that the hybrdization width decreased by a factor of order $(g /h)^2$ (the calculation is not precise enough to take the prefactor $\pi$ seriously). Iterating this and recalling that $w_0=\xi$, we get 
$$
w_i=v_{i-1}(2h_i)=\xi  (g /h)^{2i},
$$
i.e. an exponentially decaying width for the structure factors. By comparison with \ref{sec: stability one dimension}, the decay rate $\log (h/g)$ is a natural estimate for the inverse of the localization length $\zeta$. 

We recall that the whole iterative calculation makes sense only as long ETH is satisfied, hence as $v$ is smooth on the scale of the level spacing. This means that this procedure breaks down when $w_i=1/\rho_i$ where $\rho_i=2^i \rho_0\approx 2^{i+\ell_b}/\xi$, hence the breakdown is at $i=\ell_c$ with 
$$
2^{\ell_b} (2(g/h)^2)^{\ell_c} = 1
$$
which yields the same conclusion as \eqref{eq: length crossover} upon identifying $e^{-\zeta}=g/h$.

Let us now return to the validity of the FGR in this situation. The scale $\omega_0$ over which $v$ changes at $\omega$ can be estimated by $\omega_0 \approx \tfrac{v(\omega)}{\str v'(\omega)\str}$, so the FGR condition $w\ll \omega_0$ reads (for coupling the $i+1$'th spin)
$$
 g^2 \str v_i'(2h_{i+1})\str \ll 1.
$$ 
For $i=0$ this means  $g\ll \xi$, which was assumed, and for $i \geq 1$, it boils down to 
$$
(\tfrac{g}{w_i})^2 \ll   \tfrac{(1+\mathcal{M}_i^2)^2}{\mathcal{M}_i},\qquad \mathcal{M}_i\equiv \tfrac{\min{\str 2h_{i+1}\pm 2h_{i}\str}}{w_i}
$$
which is indeed satisfied in the treatment above (except at resonances).
If, however,  we were to consider more general interactions, see Section \ref{sec: general interactions}, then the FGR would typically not be satisfied when adding the $\sigma_i^z\sigma_{i+1}^z$ terms and this would lead to a more intricate theory that we do not discuss here, see however \cite{nandkishore2016general}

\section{Structure of the interface region}\label{sec: structure of the interface region}

Let us discuss the most striking properties of the spatial interface region between an ergodic and an MBL system, referred to as the crossover region in Section \ref{sec: stability against ergodic grains}.
First of all, our theory describes this interface region as fully thermal or ergodic: {\it{ETH holds for all local operators in the cross-over region}}. However, the onset of localization is revealed by the narrowing of structure factors.  Indeed, adopting the framework of Sections \ref{sec: stability one dimension} and \ref{sec: instability subexponential}, the structure factor $v_i$ of an operator located at the $i$'th added spin, has two main peaks of halfwidth $w_i$ of the order 
$$
w_i \approx g_i^{2}/\xi \approx e^{-2(i-1)/\zeta}g_1^2/\xi
$$
This  follows from applying the formula \eqref{eq: v spin} and it matches with the alternative derivation presented in Section \ref{sec: stability of mbl without}. To avoid confusion, we stress that previously the symbol $w$ was reserved for the (half)width of hyubridization functions.
Since the widths are exponentially decreasing,, we see that adding further spins (i.e.\ $i+1,i+2,\ldots$) does not significantly affect the widts of the spins already added and so $w_i$ above is the width regardless of how many more spins have been added.

To quantify the behaviour in the interface region, we calculate an \emph{inverse participation ratio} ($\ipr$) of operators $O$ in this region. By the IPR of $O$, we mean that we choose an eigenstate $\psi$ (at maximal entropy, to stay within the setup) and we look at the distribution over the other eigenstates $\psi'$ of the matrix elements
$$
 \langle \psi' \str O \str \psi \rangle
$$
The $\ipr(O)=\ipr(O,\psi)$ is then 
$$
\ipr(O) = \Big(\sum_{\psi'}   \str  \langle \psi' \str O \str \psi \rangle \str^4\Big)^{-1}
$$
Let us calculate this in terms of the parameters used in the ETH hypothesis, notably the structure factor $v$.  Then 
$$
\ipr(O) =\rho \Big(\int \d \omega      v^2(\omega)\Big)^{-1}
$$
If the structure factor $v$ consists of two peaks with halfwidth $w$, then on its support $v \approx 1/(4w) $ and this yields
$$
\ipr(O) = 4\rho w
$$
Note that hence $\ipr(O)\approx \cal N$ where $\cal N$ is the number of states or 'effective dimension' within an energy range $4w$, so this matches with the meaning of IPR in one-particle systems.  As $\cal N$ scales exponentially with system size, it is natural to consider rather the logarithm of IPR's, so we define
$$
{\cal D}(O)= \log  {\ipr}(O)
$$
Let us now apply this to the spins added to the bath, i.e.\ $O=\sigma^x_i$ for example.  Let us assume that $\ell$ spins have been added to a bath of length $\ell_b$ and these spins have been thermalized, i.e. $\ell \leq \ell_c$ with $\ell_c$ as in \eqref{eq: length crossover}. 
 For the density of states $\rho$, we of course have to use the density due to all the spins, i.e.\ $\rho_\ell=2^{\ell_b+\ell}/\xi$  (we neglect a volume denominator by pretending that $\xi$ is the total bandwidth of the bath). 
This means that we have
\begin{align}
 {\cal D}(\sigma^x_0)&= V_{th}\log{2}  \label{behavior {Ipr}}  \\
{\cal D}(\sigma^x_i)& = \log(4\rho_\ell w_i),\qquad \qquad \qquad \qquad  i\geq 1  \nonumber \\
&\sim   \log(4\rho_\ell w_1) -2(i-1)/\zeta \nonumber \\
& \sim  \log4+V_{th}\log{2}-2\log(\tfrac{g_1}{\xi})-2(i-1)/\zeta \nonumber
\end{align}
where $V_{th}= l_b+\ell$ is simply the volume of the thermal region (original bath plus thermalized spins) and we used $w_1=g_1^2/\xi$, and ${\cal D}(\sigma^x_0)=V_{th}\log{2}$ (the operator inside the bath). 

It is worth spelling out the two aspects contained in \eqref{behavior {Ipr}}, holding in fact for arbitrary operators $O_i$ located around $i$: 
First, all the spins in the cross-over region maximally participate to the effective dimension of the bath for an operator inside the ergodic grain, since $\mathcal D (O_{i=0})$ is proportional to $V_{th}$ and not to $\ell_b$. 
Second, $\mathcal D(O_i)-\mathcal D(O_1)$ decays linearly in $i$ for operators inside the cross-over region. Both of these aspects seem to be well-confirmed by the numerics in Section \ref{sec: numerical tests}.

More generally, for a subcritial grain (i.e.\@ if \eqref{size thermal volume} admits a solution for $\ell_c$) in $d\geq 1$, we can consider also $\ell \geq \ell_c$. The theory predicts a smooth transition between the core of the grain and the MBL region. The MBL region is reached when ${\cal D}(O) \approx {0}$, which of course reproduces the estimate \eqref{eq: length crossover} for $\ell_c$ in $d=1$.

\section{Numerical tests} \label{sec: numerical tests}

\subsection{LIOMs coupled to a random matrix bath}\label{sec: strenghtening by localized spins}
 In this section, we test the central prediction of our theory; that every  spin, however weakly coupled, that is thermalised by the bath, indeed doubles the effective dimension of the bath.  We consider again the Hamiltonian introduced in \eqref{eq: ham bath plus loc}:   
$$H={{{H_B}}} + \sum_{i=1}^n ( g_i \sigma_0^{x} \tau_i^{x} + h_i \tau_i^{z} )$$
where ${{{H_B}}}$ is the Hamiltonian of the bath and $\sigma^x_0$ pertains to a spin in the bath/ergodic grain at the boundary. 
The LIOM-spin $\tau_i^{z}$ is at distance $i$ (in lattice units) from the ergodic grain, hence we set
\begin{equation}\label{gs}
g_i = \alpha^{i-1} g_1 \quad (\alpha < 1).
\end{equation}
We take $\alpha = 3/4$, corresponding to inverse localization length $\zeta^{-1} = \log (4/3)$.  Since $\zeta^{-1} < \log(2)/2$, our theory predicts that the cross-over region extends to infinity: it should be able to thermalize an arbitrary amount of added LIOM-spins. 

We take $g_1 = 0.2$ and ${{h_i}} = 1 +\widetilde{h}_i$ with $\widetilde{h}_i$ i.i.d.\ random variables drawn uniformly from  $[-0.5, 0.5]$. 
These parameters should guarantee that the LIOM-spins $\tau_i$ are indeed by themselves localized degrees of freedom. This is a somewhat subtle statement because the $\tau_i$ are not coupled to each other and hence they have no mechanism for delocalization. However, we mean by this the interaction between them via the bath spin $\sigma_0^x$ is not sufficient to delocalize them. In practice, this simply means that we check (see below) whether for a sufficiently small bath, the system is localized.  
For the bath, we first consider a grain of $6$ spins with no particular spatial structure; ${{{H_B}}}$ is then simply a random matrix (RM) acting on a $2^6$-dimensional space, with level spacing $1/\rho = 0.07$ and bandwidth $4.2$.
We couple up to $8$ LIOMs to this grain. To minimize finite-size effects, we consider a bath spin operator $O_B=\sigma_B^{x}$ that is \emph{not} the one used to couple to the LIOM's, i.e.\ not  $\sigma_0^{x}$ (it does not make sense to specify $O_B$ as this bath has no spatial structure). 

\subsubsection{Strenghtening of the bath by the LIOMs}

\begin{figure}[h!]
\begin{center}
\includegraphics[width=7cm,height=4cm]{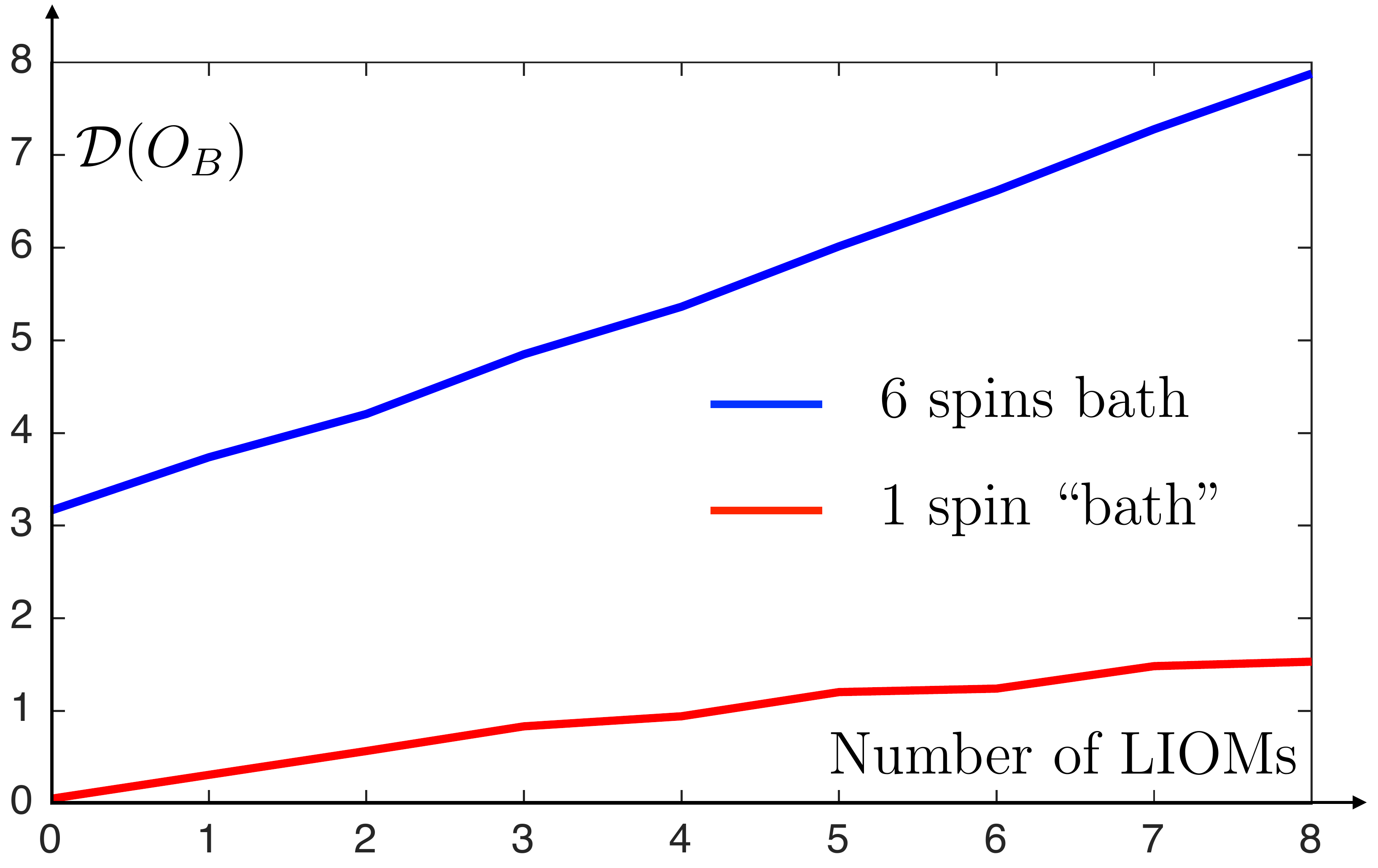}
\\[4mm]
\includegraphics[width=7cm,height=4cm]{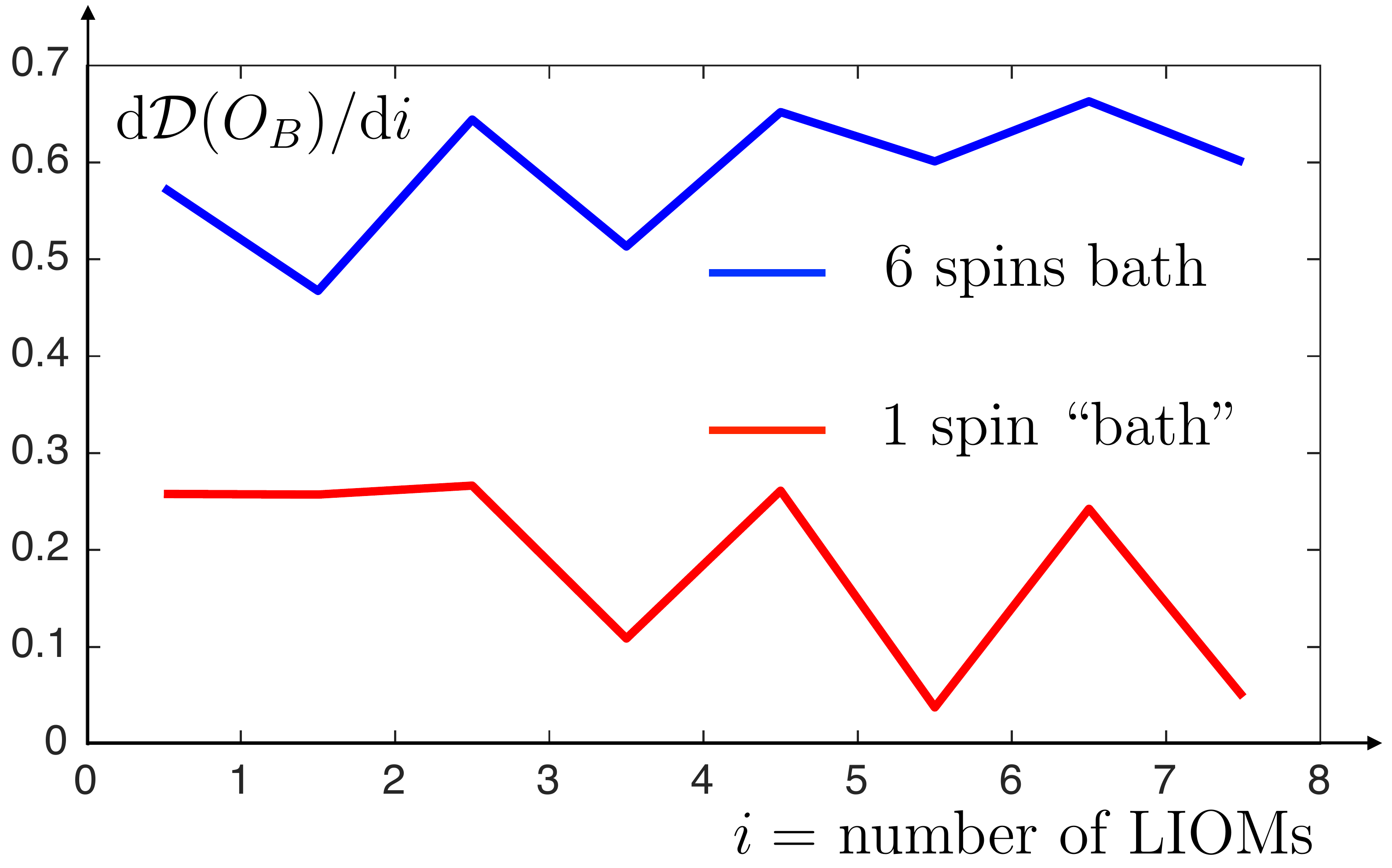}
\end{center}
\caption{
\label{fig: IPR bath}
\emph{Top}: $\mathcal D(O_{B})$ for a spin inside the bath, as a function of the number of LIOMs coupled to the bath. 
In blue, a $6$ spins RM bath,  $N=200$ ; in red, a single spin `bath', $N=4000$.
\emph{Bottom}: Discrete slopes of the above curves. At each $i+\tfrac{1}{2}$, we plot $\mathcal D(O_{B}, i+1)-\mathcal D(O_{B}, i) $, with $i$ the number of LIOMs.}
\end{figure}

We focus first on a local operator inside the core of the ergodic grain.
We compute the disorder-average of $\mathcal D(O_B)$, as a function of the number of LIOMs coupled to the bath.
The result is represented by the blue line on the left panel of figure  \ref{fig: IPR bath}, while the corresponding slope of this curve is depicted on the right panel.
The value of this slope approaches the ideal value $\log 2$ predicted by our theory. 

It remains to check that the spins $\tau_i$ are localized in the absence of the bath. 
For this, we replace the $6$ spins bath, by a single spin ('bandwidth'=2.25), for which perturbative computations predict the persistence of localization. 
We still compute the disorder-average of $\mathcal D(O_B)$, where $O_B$ corresponds now to the unique spin in the `bath'. 
The result is represented by the red lines on figure  \ref{fig: IPR bath}. 
The slope much smaller than in the previous case (less than half of the ideal value) and we observe a tendency for the slope to decay.

\subsubsection{The cross-over region}\label{sec: crossover region}

\begin{figure}[h!]
\begin{center}

\includegraphics[width=8.5cm,height=5cm]{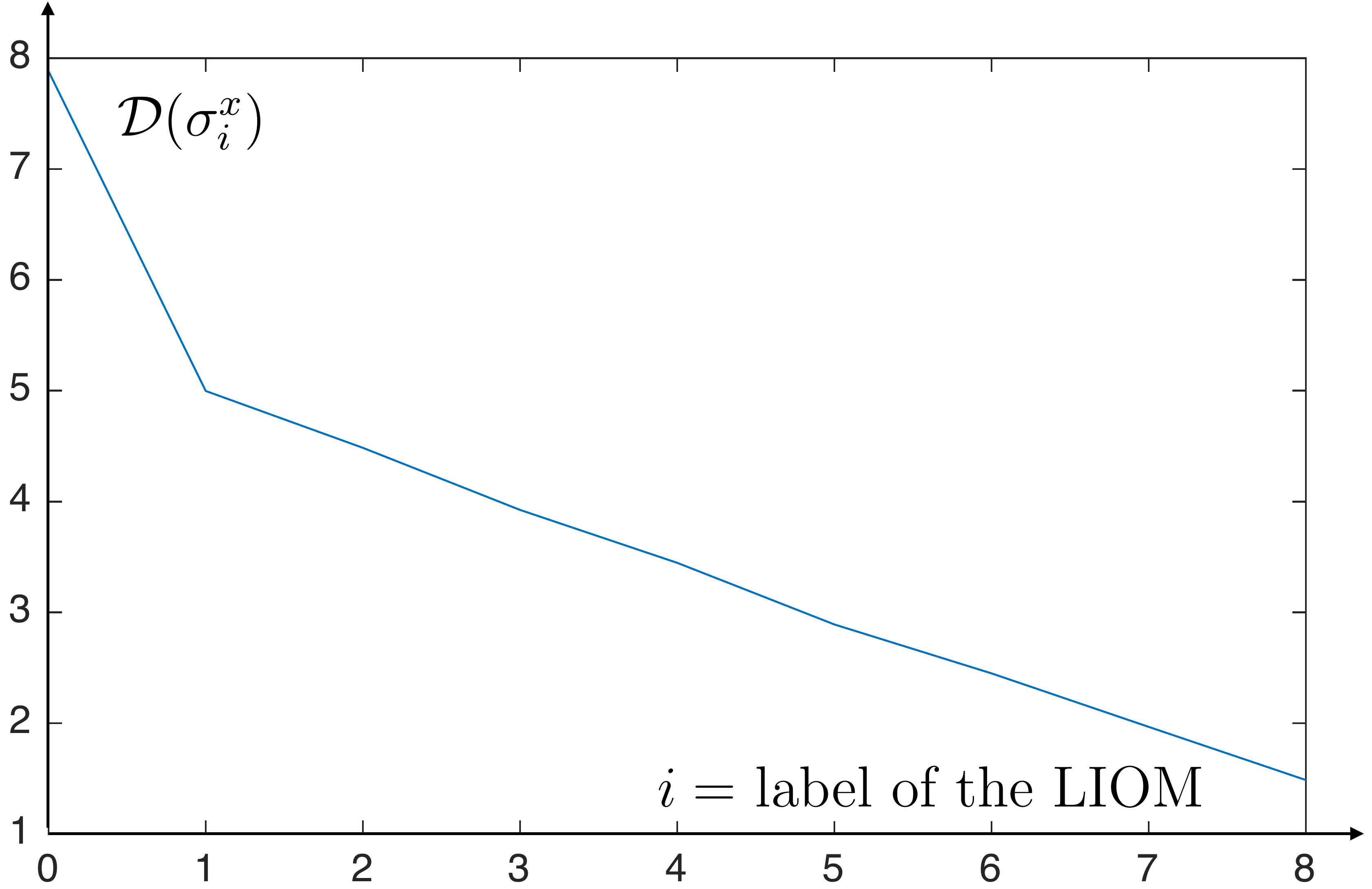}

\end{center}
\caption{
\label{fig: IPR buffer region} 
$\mathcal D(\sigma_i^{x})$ as a function of $i$ for $0 \le i \le 8$, for a system of 8 localized spins coupled to a $6$ spins RM bath. $N=200$ for each point of the curve}

\end{figure}

We now turn to the investigation of the spatial dependence of $\log \ipr$ in the cross-over region. 
Despite notational similarity, the quantity we consider is different from that on Figure \eqref{fig: IPR bath}, where $O$ was a bath operator and the dependence on $i$ was simply because of the number of attached LIOM's. 
We now calculate $\mathcal D(\sigma_i^{x})$ for all $0 \le i \le 8$ (where $i=0$ corresponds to a spin inside the bath). 
Our theory, see Section \ref{sec: structure of the interface region}, predicts that $\mathcal D(O_i)-\mathcal D(O_1), i\geq 1$ decays linearly with slope $2(\log 4 - \log 3) = 0.58$, wheres the decay from $D(O_0)$ to $D(O_1)$ is unrelated.
The actual result of the calculation is depicted on figure \ref{fig: IPR buffer region}. 
We indeed observe a linear decay from spin $1$ to $8$. The slope in the linear region is $0.5$, which seems in reasonable agreement with the theoretical value $0.58$.

\subsection{A universal thermalization curve: comparison of different baths}\label{sec: universal thermalization curve}

In this section, we compare the action of different baths, differing in size and nature: random matrix, local Hamiltonian, local Hamiltonian plus thermalized LIOM's. Already the fact that the latter can be labeled a 'bath' is a nontrivial element and, in some sense, the core of the message of this paper. 
  In our theory, the characteristics of the bath enter only via the dimensionless coupling constant $\mathcal G$ and via the hybridization width $w$ describing the thermalization of a spin. Moreover, if the FGR holds, then $w$ is rigidly related to $\mathcal G$ by $\mathcal{G}^2=\rho w $, hence we focus on the dependence on $\mathcal{G}$ here.  

\begin{figure}[h!]
\begin{center}
\includegraphics[width=8.5cm,height=5cm]{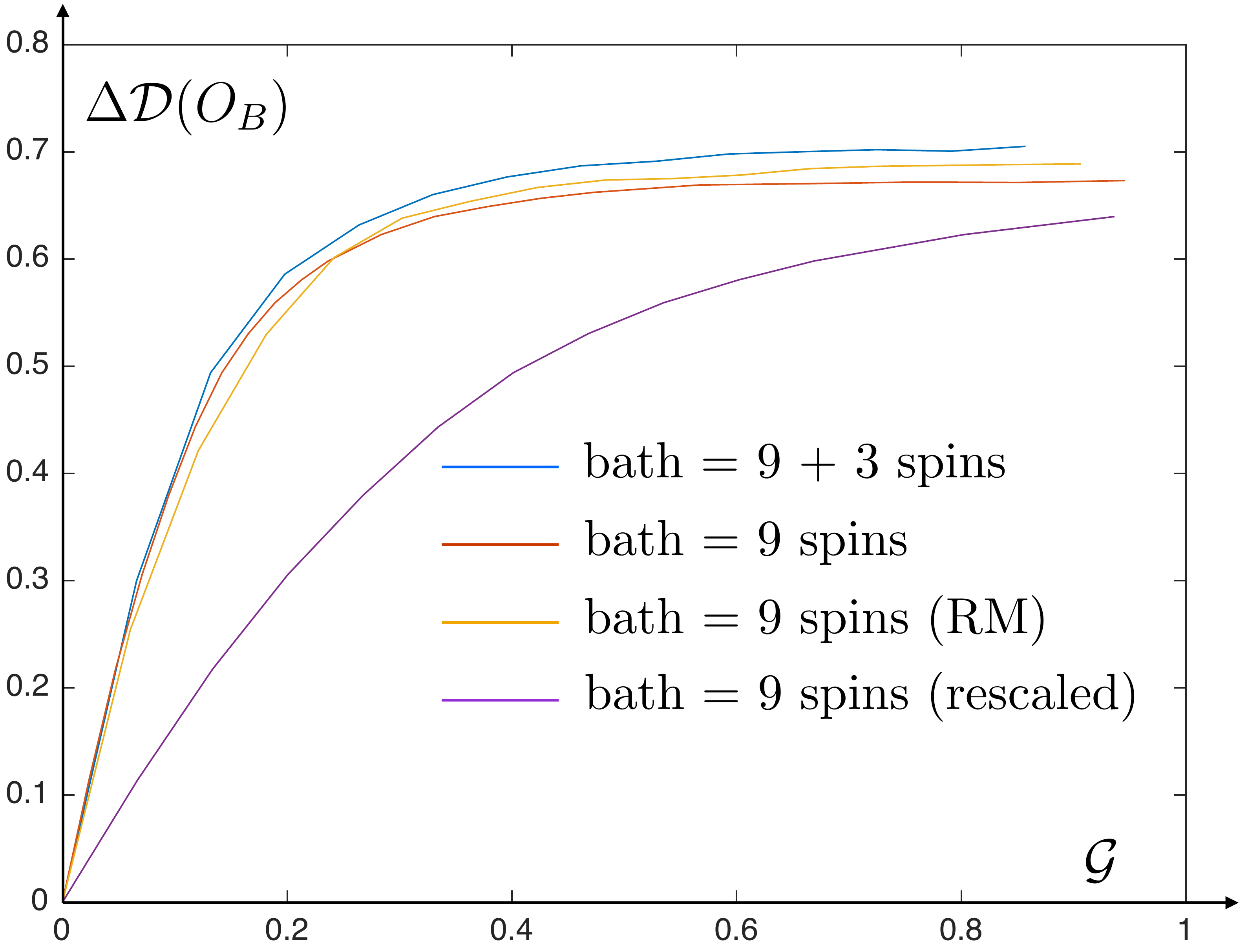}

\end{center}
\caption{
\label{fig: IPR z}
$\Delta \mathcal{D}(O_B)=\mathcal D(O_B)\big\str_{\cal G}-\mathcal D(O_B)\big\str_{{\cal G}=0}$ as a function of $\mathcal G$. 
1)\textcolor{blue}{blue}: $B'$ = $9$ spins bath with local interactions and $3$ thermalized LIOMs, $N=200$ (number of disorder realizations); 
2)\textcolor{RedOrange}{Orange}: $B'$ = $9$ spins bath with local interactions, $N=1000$; 
3) \textcolor{Dandelion}{yellow}: $B'$ = $9$ spin RM bath, $N=100$, 
4) \textcolor{Purple}{purple}: $B'$ = $9$ spins bath with local interactions and $3$ un-thermalized LIOMs.}
\end{figure}

The $\mathcal{G}$ parameter enters our theory in deciding whether or not an external spin is thermalized. However, numerics shows that $\mathcal{G}$ also accurately predicts the quality of the thermalization.
This quality of thermalization is quantified by the increase in $\mathcal D(O_B)$ upon adding the spin, e.g.\ 
$$ \Delta \mathcal{D}(O_B) :=\mathcal D(O_B)\big\str_{\cal G}-\mathcal D(O_B)\big\str_{{\cal G}=0} $$ as a function of $\mathcal G$.
Of course, $\Delta \mathcal{D}(O_B)=0$ for $\mathcal{G}=0$ and one expects that the effective dimension doubles by inclusion of the spin, hence $\Delta \mathcal{D}(O_B) \to \log 2$  as $\mathcal{G}\gg 1$. 
The behavour for intermediate value of $\mathcal{G}$ is in practice not so important for our theory.  However, it is remarkable
 that the curves in Figure \ref{fig: IPR z} collapse rather well for different baths (the deviant lowest curve is explained below) and we feel it supports our theory in a compelling way.

Let us now discuss in detail the 3 different baths that we compare: 

\noindent \emph{\textcolor{blue}{Local $9$-spin bath plus 3 LIOMs}}  We introduce a  bath of $9$ spins with a local structure: 
\begin{equation}\label{Hamiltonian bath}
{{{H_B}}} = \sum_{i=1}^{9} (h'_i \sigma_i^{z} + J \sigma_i^{x}) + J'\sum_{i=1}^{8} \sigma_i^{x}\sigma_{i+1}^{x}
\end{equation}
with $h_i' = 1 + \widetilde{h}_i'$ with $-0.3 < \widetilde{h}_i' < 0.3$, $J=0.5$ and $J'=1$. 
We couple $4$ LIOMs $\tau_{1,\ldots,4}$ (we take $h_i$, $g_1=0.2$ and $\alpha=3/4$ as in Section \ref{sec: strenghtening by localized spins}) to this bath, but we vary the coupling strength $g_4$ of the last one. So, the 9-spin bath plus $3$ LIOMs form a bath, let us call it $B'$, that we investigate
by viewing how it acts on the $4$th spin $\tau_4$. As explained above, this is done by by measuring $\mathcal D(O_B)$ for an operator $O_B$ in the original bath $B$. We choose $O_B=\sigma^x_4$ (this '4' refers to a site in the bath and is completely unrelated to the $4$ of $\tau_4$).  Note that $\sigma^x_9$ is the same as $\sigma^x_0$ in Section \ref{sec: strenghtening by localized spins}.

We determine $\cal G$ for the coupling of the fourth spin to $B'$:  
${\cal G}=g_4 \sqrt{v(2h_4)\rho}$ with $1/\rho$ the level spacing in $B'$ and $v(\cdot)$ the structure factor of the bath operator $\sigma_0^x$. 
Numerically, we calculated $\cal G$ as (the disorder average of) $g_4 \rho \sqrt{M}$
where
$$
M:= \frac{1}{\cal N} \sum_{b',s': \str E(b',s')-E(b,s) \str \leq 1}  \delta_{s,-s'} \str\langle b\str \sigma_0^x \str b' \rangle \str^2
$$
with $\cal N$ the number of terms in the sum, $b,b'$ eigenstates of the bath $B'$, $s$ an eigenstates of $\tau^z_4$ and $b,s$ chosen so that $\str b,s\rangle$ eigenstate with energy $E(b,s)$ closest to $0$ (maximal entropy).   

\noindent \emph{\textcolor{RedOrange}{Local $9$-spin bath}} Here we take the same $9$-spin bath as above but without the 3 LIOMs coupled to it. 

\noindent \emph{\textcolor{Dandelion}{Random matrix}}
This is a random matrix bath mimicking the 9-spin bath (with dimension $2^9$ and level spacing $1/\rho = 0.018$), no extra localized spins.

For contrast, we plot an example of a curve that doesn't match. 
The \textcolor{Purple}{purple curve} on figure \ref{fig: IPR z} is built from the same data as the orange curve, but is scaled differently.  
Indeed, one considers here the case where there are actually three intermediate spins, as for the blue curve, but where these spins are uncoupled $(g_1 = g_2 = g_3 = 0)$. 
It is thus clear that they have no physical effect, so that the data are the same as the one obtained on the orange curve; however, if one considers the original bath together with these three spins as the new bath $B'$, 
one divides the level spacing by $2^3$, leading to a different definition of $ \mathcal G$ w.r.t.\@ the orange curve. 
We see that the purple curve is clearly an outlier, indicating that when the three intermediate spins are truly coupled, they do participate to the thermalization of the last $4$th spin.

Finally,  we observe that the value of $\Delta \mathcal D$ for the blue curve  becomes even slightly larger than $\log 2$ for values of $\mathcal G$ close to 1. 
This is a priori surprising as the increase of dimension from a single spin should be $\log 2$ at most.  We suspect the following scenario:  before coupling the $4$th spin, $\mathcal D$ did not reach its maximal value, because the three coupled spins were only imperfectly thermalized. Indeed, in our numerics the bare values used for the first 3 spins are not so far from the critical value needed for thermalization. The first of these spins corresponds to ${\cal G}= 0.67$, which can still be considered slightly in the transition region, as is revealed by a careful inspection of the curves of Figure \ref{fig: IPR z}. However, the situation becomes better for each added spin (this is a direct consequence of the fact that $\log (4/3)>\log{2}/2$ and there is catchup effect: the fourth spin helps to complete the imperfect thermalization of the first three spins.

\section{Conclusion}\label{sec: conclusion}
{
We have proposed a simple RMT theory for the joint eigenfunctions of spins coupled to a finite bath.
When we apply this theory repeatedly, we obtain a clear-cut prediction whether the system will be localized or ergodic. 
In this way, we investigate finite ergodic grain (bath) coupled to well-localized spins, i.e.\ with coupling strength small compared to disorder. 
The predictions are consistent with localization in $d=1$ with exponentially decaying interactions:  localization is stable with respect to such ergodic grains. For interactions that decay slower, or in $d>1$, our theory predicts delocalization. 
The most quantitative result of our theory is a description of the spatial intermediate region between an ergodic grain and a $d=1$ localized material. In particular, it states that any degree of freedom, say a $\tfrac{1}{2}$-spin, in this region enhances the ergodic grain, leading in particular to a doubling of the effective dimension or the IPR (inverse participation ratio) parameter. This prediction is reasonably validated by numerics, though there is need for finer tests. As an aside, the numerics reveals a rather universal pattern of thermalization that does not distinghuish between the action of a bath on very localized degrees of freedom or others, i.e.\ (nearly) ergodic degrees of freedom.  
 }

\begin{acknowledgments}
 This paper originates in discussions with M. Mueller, J. Imbrie, S. Gopalakrishnan and S. Parameswaran. In a later stage, we profited a lot from comments by E. Altman and E. Berg on an early version of this work. We are very grateful for all these interactions. 
 The help of M. Serbyn with the numerics is also much appreciated.
  W.D.R.\@  acknowledges the grant from the Deutsche Forschungsgesellschaft, DFG No. RO 4522/1-1. 
Both F.H.\@ and W.D.R.\@ acknowledge the support of the ANR grant JCJC, and thank the CNRS InPhyNiTi Grant (MaBoLo) for financial support.
\end{acknowledgments}


\bibliographystyle{rsc}

\bibliography{loclibrary}

\providecommand*{\mcitethebibliography}{\thebibliography}
\csname @ifundefined\endcsname{endmcitethebibliography}
{\let\endmcitethebibliography\endthebibliography}{}
\begin{mcitethebibliography}{48}
\providecommand*{\natexlab}[1]{#1}
\providecommand*{\mciteSetBstSublistMode}[1]{}
\providecommand*{\mciteSetBstMaxWidthForm}[2]{}
\providecommand*{\mciteBstWouldAddEndPuncttrue}
  {\def\EndOfBibitem{\unskip.}}
\providecommand*{\mciteBstWouldAddEndPunctfalse}
  {\let\EndOfBibitem\relax}
\providecommand*{\mciteSetBstMidEndSepPunct}[3]{}
\providecommand*{\mciteSetBstSublistLabelBeginEnd}[3]{}
\providecommand*{\EndOfBibitem}{}
\mciteSetBstSublistMode{f}
\mciteSetBstMaxWidthForm{subitem}
{(\emph{\alph{mcitesubitemcount}})}
\mciteSetBstSublistLabelBeginEnd{\mcitemaxwidthsubitemform\space}
{\relax}{\relax}

\bibitem[Anderson(1958)]{anderson1958absence}
P.~W. Anderson, \emph{Physical review}, 1958, \textbf{109}, 1492\relax
\mciteBstWouldAddEndPuncttrue
\mciteSetBstMidEndSepPunct{\mcitedefaultmidpunct}
{\mcitedefaultendpunct}{\mcitedefaultseppunct}\relax
\EndOfBibitem
\bibitem[Basko \emph{et~al.}(2006)Basko, Aleiner, and
  Altshuler]{basko2006metal}
D.~Basko, I.~Aleiner and B.~Altshuler, \emph{Annals of physics}, 2006,
  \textbf{321}, 1126--1205\relax
\mciteBstWouldAddEndPuncttrue
\mciteSetBstMidEndSepPunct{\mcitedefaultmidpunct}
{\mcitedefaultendpunct}{\mcitedefaultseppunct}\relax
\EndOfBibitem
\bibitem[Gornyi \emph{et~al.}(2005)Gornyi, Mirlin, and
  Polyakov]{gornyi2005interacting}
I.~Gornyi, A.~Mirlin and D.~Polyakov, \emph{Physical review letters}, 2005,
  \textbf{95}, 206603\relax
\mciteBstWouldAddEndPuncttrue
\mciteSetBstMidEndSepPunct{\mcitedefaultmidpunct}
{\mcitedefaultendpunct}{\mcitedefaultseppunct}\relax
\EndOfBibitem
\bibitem[Oganesyan and Huse(2007)]{oganesyan2007localization}
V.~Oganesyan and D.~A. Huse, \emph{Physical Review B}, 2007, \textbf{75},
  155111\relax
\mciteBstWouldAddEndPuncttrue
\mciteSetBstMidEndSepPunct{\mcitedefaultmidpunct}
{\mcitedefaultendpunct}{\mcitedefaultseppunct}\relax
\EndOfBibitem
\bibitem[Pal and Huse(2010)]{pal2010many}
A.~Pal and D.~A. Huse, \emph{Physical review b}, 2010, \textbf{82},
  174411\relax
\mciteBstWouldAddEndPuncttrue
\mciteSetBstMidEndSepPunct{\mcitedefaultmidpunct}
{\mcitedefaultendpunct}{\mcitedefaultseppunct}\relax
\EndOfBibitem
\bibitem[Khemani \emph{et~al.}(2016)Khemani, Pollmann, and
  Sondhi]{khemani2016obtaining}
V.~Khemani, F.~Pollmann and S.~Sondhi, \emph{Physical Review Letters}, 2016,
  \textbf{116}, 247204\relax
\mciteBstWouldAddEndPuncttrue
\mciteSetBstMidEndSepPunct{\mcitedefaultmidpunct}
{\mcitedefaultendpunct}{\mcitedefaultseppunct}\relax
\EndOfBibitem
\bibitem[Kj{\"a}ll \emph{et~al.}(2014)Kj{\"a}ll, Bardarson, and
  Pollmann]{kjall2014many}
J.~A. Kj{\"a}ll, J.~H. Bardarson and F.~Pollmann, \emph{Physical review
  letters}, 2014, \textbf{113}, 107204\relax
\mciteBstWouldAddEndPuncttrue
\mciteSetBstMidEndSepPunct{\mcitedefaultmidpunct}
{\mcitedefaultendpunct}{\mcitedefaultseppunct}\relax
\EndOfBibitem
\bibitem[Luitz \emph{et~al.}(2015)Luitz, Laflorencie, and Alet]{luitz2015many}
D.~J. Luitz, N.~Laflorencie and F.~Alet, \emph{Physical Review B}, 2015,
  \textbf{91}, 081103\relax
\mciteBstWouldAddEndPuncttrue
\mciteSetBstMidEndSepPunct{\mcitedefaultmidpunct}
{\mcitedefaultendpunct}{\mcitedefaultseppunct}\relax
\EndOfBibitem
\bibitem[Schreiber \emph{et~al.}(2015)Schreiber, Hodgman, Bordia, L{\"u}schen,
  Fischer, Vosk, Altman, Schneider, and Bloch]{schreiber2015observation}
M.~Schreiber, S.~S. Hodgman, P.~Bordia, H.~P. L{\"u}schen, M.~H. Fischer,
  R.~Vosk, E.~Altman, U.~Schneider and I.~Bloch, \emph{Science}, 2015,
  \textbf{349}, 842--845\relax
\mciteBstWouldAddEndPuncttrue
\mciteSetBstMidEndSepPunct{\mcitedefaultmidpunct}
{\mcitedefaultendpunct}{\mcitedefaultseppunct}\relax
\EndOfBibitem
\bibitem[Smith \emph{et~al.}(2015)Smith, Lee, Richerme, Neyenhuis, Hess, Hauke,
  Heyl, Huse, and Monroe]{smith2015many}
J.~Smith, A.~Lee, P.~Richerme, B.~Neyenhuis, P.~W. Hess, P.~Hauke, M.~Heyl,
  D.~A. Huse and C.~Monroe, \emph{arXiv preprint arXiv:1508.07026}, 2015\relax
\mciteBstWouldAddEndPuncttrue
\mciteSetBstMidEndSepPunct{\mcitedefaultmidpunct}
{\mcitedefaultendpunct}{\mcitedefaultseppunct}\relax
\EndOfBibitem
\bibitem[Serbyn \emph{et~al.}(2013)Serbyn, Papi{\'c}, and
  Abanin]{serbyn2013local}
M.~Serbyn, Z.~Papi{\'c} and D.~A. Abanin, \emph{Physical review letters}, 2013,
  \textbf{111}, 127201\relax
\mciteBstWouldAddEndPuncttrue
\mciteSetBstMidEndSepPunct{\mcitedefaultmidpunct}
{\mcitedefaultendpunct}{\mcitedefaultseppunct}\relax
\EndOfBibitem
\bibitem[Imbrie(2016)]{imbrie2016many}
J.~Z. Imbrie, \emph{Journal of Statistical Physics}, 2016, \textbf{163},
  998--1048\relax
\mciteBstWouldAddEndPuncttrue
\mciteSetBstMidEndSepPunct{\mcitedefaultmidpunct}
{\mcitedefaultendpunct}{\mcitedefaultseppunct}\relax
\EndOfBibitem
\bibitem[Huse \emph{et~al.}(2014)Huse, Nandkishore, and
  Oganesyan]{huse2014phenomenology}
D.~A. Huse, R.~Nandkishore and V.~Oganesyan, \emph{Physical Review B}, 2014,
  \textbf{90}, 174202\relax
\mciteBstWouldAddEndPuncttrue
\mciteSetBstMidEndSepPunct{\mcitedefaultmidpunct}
{\mcitedefaultendpunct}{\mcitedefaultseppunct}\relax
\EndOfBibitem
\bibitem[Note1()]{Note1}
Although, in our opinion \cite {de2016absence}, systems with a mobility edge
  are also 'merely' MBL-like or glassy.\relax
\mciteBstWouldAddEndPunctfalse
\mciteSetBstMidEndSepPunct{\mcitedefaultmidpunct}
{}{\mcitedefaultseppunct}\relax
\EndOfBibitem
\bibitem[Basko(2011)]{basko2011weak}
D.~Basko, \emph{Annals of Physics}, 2011, \textbf{326}, 1577--1655\relax
\mciteBstWouldAddEndPuncttrue
\mciteSetBstMidEndSepPunct{\mcitedefaultmidpunct}
{\mcitedefaultendpunct}{\mcitedefaultseppunct}\relax
\EndOfBibitem
\bibitem[Huveneers(2013)]{huveneers2013drastic}
F.~Huveneers, \emph{Nonlinearity}, 2013, \textbf{26}, 837\relax
\mciteBstWouldAddEndPuncttrue
\mciteSetBstMidEndSepPunct{\mcitedefaultmidpunct}
{\mcitedefaultendpunct}{\mcitedefaultseppunct}\relax
\EndOfBibitem
\bibitem[Deutsch(1991)]{deutsch1991quantum}
J.~Deutsch, \emph{Physical Review A}, 1991, \textbf{43}, 2046\relax
\mciteBstWouldAddEndPuncttrue
\mciteSetBstMidEndSepPunct{\mcitedefaultmidpunct}
{\mcitedefaultendpunct}{\mcitedefaultseppunct}\relax
\EndOfBibitem
\bibitem[Srednicki(1994)]{srednicki1994chaos}
M.~Srednicki, \emph{Physical Review E}, 1994, \textbf{50}, 888\relax
\mciteBstWouldAddEndPuncttrue
\mciteSetBstMidEndSepPunct{\mcitedefaultmidpunct}
{\mcitedefaultendpunct}{\mcitedefaultseppunct}\relax
\EndOfBibitem
\bibitem[Potter \emph{et~al.}(2015)Potter, Vasseur, and
  Parameswaran]{potter2015universal}
A.~C. Potter, R.~Vasseur and S.~Parameswaran, \emph{Physical Review X}, 2015,
  \textbf{5}, 031033\relax
\mciteBstWouldAddEndPuncttrue
\mciteSetBstMidEndSepPunct{\mcitedefaultmidpunct}
{\mcitedefaultendpunct}{\mcitedefaultseppunct}\relax
\EndOfBibitem
\bibitem[Vosk \emph{et~al.}(2015)Vosk, Huse, and Altman]{vosk2015theory}
R.~Vosk, D.~A. Huse and E.~Altman, \emph{Physical Review X}, 2015, \textbf{5},
  031032\relax
\mciteBstWouldAddEndPuncttrue
\mciteSetBstMidEndSepPunct{\mcitedefaultmidpunct}
{\mcitedefaultendpunct}{\mcitedefaultseppunct}\relax
\EndOfBibitem
\bibitem[Zhang \emph{et~al.}(2016)Zhang, Zhao, Devakul, and
  Huse]{zhang2016many}
L.~Zhang, B.~Zhao, T.~Devakul and D.~A. Huse, \emph{Physical Review B}, 2016,
  \textbf{93}, 224201\relax
\mciteBstWouldAddEndPuncttrue
\mciteSetBstMidEndSepPunct{\mcitedefaultmidpunct}
{\mcitedefaultendpunct}{\mcitedefaultseppunct}\relax
\EndOfBibitem
\bibitem[Vonnegut(1963)]{vonnegut1963cradle}
K.~Vonnegut, \emph{Cat's Cradle}, Holt, Rinehart and Winston, 1963\relax
\mciteBstWouldAddEndPuncttrue
\mciteSetBstMidEndSepPunct{\mcitedefaultmidpunct}
{\mcitedefaultendpunct}{\mcitedefaultseppunct}\relax
\EndOfBibitem
\bibitem[Abanin \emph{et~al.}(2015)Abanin, De~Roeck, Huveneers, and
  Ho]{abanin2015asymptotic}
D.~Abanin, W.~De~Roeck, F.~Huveneers and W.~W. Ho, \emph{arXiv preprint
  arXiv:1509.05386}, 2015\relax
\mciteBstWouldAddEndPuncttrue
\mciteSetBstMidEndSepPunct{\mcitedefaultmidpunct}
{\mcitedefaultendpunct}{\mcitedefaultseppunct}\relax
\EndOfBibitem
\bibitem[Parameswaran and Gopalakrishnan(2016)]{parameswaran2016spin}
S.~Parameswaran and S.~Gopalakrishnan, \emph{arXiv preprint arXiv:1603.08933},
  2016\relax
\mciteBstWouldAddEndPuncttrue
\mciteSetBstMidEndSepPunct{\mcitedefaultmidpunct}
{\mcitedefaultendpunct}{\mcitedefaultseppunct}\relax
\EndOfBibitem
\bibitem[Aizenman and Warzel(2011)]{aizenman2011resonant}
M.~Aizenman and S.~Warzel, \emph{arXiv preprint arXiv:1104.0969}, 2011\relax
\mciteBstWouldAddEndPuncttrue
\mciteSetBstMidEndSepPunct{\mcitedefaultmidpunct}
{\mcitedefaultendpunct}{\mcitedefaultseppunct}\relax
\EndOfBibitem
\bibitem[Chandran \emph{et~al.}(2016)Chandran, Pal, Laumann, and
  Scardicchio]{chandran2016many}
A.~Chandran, A.~Pal, C.~Laumann and A.~Scardicchio, \emph{arXiv preprint
  arXiv:1605.00655}, 2016\relax
\mciteBstWouldAddEndPuncttrue
\mciteSetBstMidEndSepPunct{\mcitedefaultmidpunct}
{\mcitedefaultendpunct}{\mcitedefaultseppunct}\relax
\EndOfBibitem
\bibitem[Agarwal \emph{et~al.}(2015)Agarwal, Gopalakrishnan, Knap, M{\"u}ller,
  and Demler]{agarwal2015anomalous}
K.~Agarwal, S.~Gopalakrishnan, M.~Knap, M.~M{\"u}ller and E.~Demler,
  \emph{Physical review letters}, 2015, \textbf{114}, 160401\relax
\mciteBstWouldAddEndPuncttrue
\mciteSetBstMidEndSepPunct{\mcitedefaultmidpunct}
{\mcitedefaultendpunct}{\mcitedefaultseppunct}\relax
\EndOfBibitem
\bibitem[Nandkishore \emph{et~al.}(2014)Nandkishore, Gopalakrishnan, and
  Huse]{nandkishore2014spectral}
R.~Nandkishore, S.~Gopalakrishnan and D.~A. Huse, \emph{Physical Review B},
  2014, \textbf{90}, 064203\relax
\mciteBstWouldAddEndPuncttrue
\mciteSetBstMidEndSepPunct{\mcitedefaultmidpunct}
{\mcitedefaultendpunct}{\mcitedefaultseppunct}\relax
\EndOfBibitem
\bibitem[Johri \emph{et~al.}(2015)Johri, Nandkishore, and Bhatt]{johri2015many}
S.~Johri, R.~Nandkishore and R.~Bhatt, \emph{Physical review letters}, 2015,
  \textbf{114}, 117401\relax
\mciteBstWouldAddEndPuncttrue
\mciteSetBstMidEndSepPunct{\mcitedefaultmidpunct}
{\mcitedefaultendpunct}{\mcitedefaultseppunct}\relax
\EndOfBibitem
\bibitem[Levi \emph{et~al.}(2015)Levi, Heyl, Lesanovsky, and
  Garrahan]{levi2015survives}
E.~Levi, M.~Heyl, I.~Lesanovsky and J.~P. Garrahan, \emph{arXiv preprint
  arXiv:1510.04634}, 2015\relax
\mciteBstWouldAddEndPuncttrue
\mciteSetBstMidEndSepPunct{\mcitedefaultmidpunct}
{\mcitedefaultendpunct}{\mcitedefaultseppunct}\relax
\EndOfBibitem
\bibitem[Fischer \emph{et~al.}(2016)Fischer, Maksymenko, and
  Altman]{fischer2016dynamics}
M.~H. Fischer, M.~Maksymenko and E.~Altman, \emph{Physical review letters},
  2016, \textbf{116}, 160401\relax
\mciteBstWouldAddEndPuncttrue
\mciteSetBstMidEndSepPunct{\mcitedefaultmidpunct}
{\mcitedefaultendpunct}{\mcitedefaultseppunct}\relax
\EndOfBibitem
\bibitem[Mukerjee \emph{et~al.}(2006)Mukerjee, Oganesyan, and
  Huse]{mukerjee2006statistical}
S.~Mukerjee, V.~Oganesyan and D.~Huse, \emph{Physical Review B}, 2006,
  \textbf{73}, 035113\relax
\mciteBstWouldAddEndPuncttrue
\mciteSetBstMidEndSepPunct{\mcitedefaultmidpunct}
{\mcitedefaultendpunct}{\mcitedefaultseppunct}\relax
\EndOfBibitem
\bibitem[Abanin \emph{et~al.}(2015)Abanin, De~Roeck, and
  Huveneers]{abanin2015exponentially}
D.~A. Abanin, W.~De~Roeck and F.~Huveneers, \emph{Physical review letters},
  2015, \textbf{115}, 256803\relax
\mciteBstWouldAddEndPuncttrue
\mciteSetBstMidEndSepPunct{\mcitedefaultmidpunct}
{\mcitedefaultendpunct}{\mcitedefaultseppunct}\relax
\EndOfBibitem
\bibitem[D'Alessio \emph{et~al.}(2015)D'Alessio, Kafri, Polkovnikov, and
  Rigol]{d2015quantum}
L.~D'Alessio, Y.~Kafri, A.~Polkovnikov and M.~Rigol, \emph{arXiv preprint
  arXiv:1509.06411}, 2015\relax
\mciteBstWouldAddEndPuncttrue
\mciteSetBstMidEndSepPunct{\mcitedefaultmidpunct}
{\mcitedefaultendpunct}{\mcitedefaultseppunct}\relax
\EndOfBibitem
\bibitem[Rigol \emph{et~al.}(2008)Rigol, Dunjko, and
  Olshanii]{rigol2008thermalization}
M.~Rigol, V.~Dunjko and M.~Olshanii, \emph{Nature}, 2008, \textbf{452},
  854--858\relax
\mciteBstWouldAddEndPuncttrue
\mciteSetBstMidEndSepPunct{\mcitedefaultmidpunct}
{\mcitedefaultendpunct}{\mcitedefaultseppunct}\relax
\EndOfBibitem
\bibitem[Khatami \emph{et~al.}(2013)Khatami, Pupillo, Srednicki, and
  Rigol]{khatami2013fluctuation}
E.~Khatami, G.~Pupillo, M.~Srednicki and M.~Rigol, \emph{Physical review
  letters}, 2013, \textbf{111}, 050403\relax
\mciteBstWouldAddEndPuncttrue
\mciteSetBstMidEndSepPunct{\mcitedefaultmidpunct}
{\mcitedefaultendpunct}{\mcitedefaultseppunct}\relax
\EndOfBibitem
\bibitem[Note2()]{Note2}
Up to corrections vanishing in the thermodynamic limit\relax
\mciteBstWouldAddEndPuncttrue
\mciteSetBstMidEndSepPunct{\mcitedefaultmidpunct}
{\mcitedefaultendpunct}{\mcitedefaultseppunct}\relax
\EndOfBibitem
\bibitem[Altman and Berg()]{altmanbergprivate}
E.~Altman and E.~Berg, \emph{private communication}\relax
\mciteBstWouldAddEndPuncttrue
\mciteSetBstMidEndSepPunct{\mcitedefaultmidpunct}
{\mcitedefaultendpunct}{\mcitedefaultseppunct}\relax
\EndOfBibitem
\bibitem[Note3()]{Note3}
This correction is not sufficient to solve the problems addressed below, and
  will be ignored here.\relax
\mciteBstWouldAddEndPunctfalse
\mciteSetBstMidEndSepPunct{\mcitedefaultmidpunct}
{}{\mcitedefaultseppunct}\relax
\EndOfBibitem
\bibitem[Berry(1977)]{berry1977regular}
M.~V. Berry, \emph{Journal of Physics A: Mathematical and General}, 1977,
  \textbf{10}, 2083\relax
\mciteBstWouldAddEndPuncttrue
\mciteSetBstMidEndSepPunct{\mcitedefaultmidpunct}
{\mcitedefaultendpunct}{\mcitedefaultseppunct}\relax
\EndOfBibitem
\bibitem[Hyatt \emph{et~al.}(2016)Hyatt, Garrison, Potter, and
  Bauer]{hyatt2016many}
K.~Hyatt, J.~R. Garrison, A.~C. Potter and B.~Bauer, \emph{arXiv preprint
  arXiv:1601.07184}, 2016\relax
\mciteBstWouldAddEndPuncttrue
\mciteSetBstMidEndSepPunct{\mcitedefaultmidpunct}
{\mcitedefaultendpunct}{\mcitedefaultseppunct}\relax
\EndOfBibitem
\bibitem[Nandkishore(2015)]{nandkishore2015many}
R.~Nandkishore, \emph{Physical Review B}, 2015, \textbf{92}, 245141\relax
\mciteBstWouldAddEndPuncttrue
\mciteSetBstMidEndSepPunct{\mcitedefaultmidpunct}
{\mcitedefaultendpunct}{\mcitedefaultseppunct}\relax
\EndOfBibitem
\bibitem[Note4()]{Note4}
The more precise statement should be that terms that affect only two LIOM's are
  subexponentially decaying in the distance. Terms that affect $n$ LIOMs decay
  in addition exponentially in $n$\relax
\mciteBstWouldAddEndPuncttrue
\mciteSetBstMidEndSepPunct{\mcitedefaultmidpunct}
{\mcitedefaultendpunct}{\mcitedefaultseppunct}\relax
\EndOfBibitem
\bibitem[Burin(2015)]{burin2015many}
A.~L. Burin, \emph{Physical Review B}, 2015, \textbf{91}, 094202\relax
\mciteBstWouldAddEndPuncttrue
\mciteSetBstMidEndSepPunct{\mcitedefaultmidpunct}
{\mcitedefaultendpunct}{\mcitedefaultseppunct}\relax
\EndOfBibitem
\bibitem[Hauke and Heyl(2015)]{hauke2015many}
P.~Hauke and M.~Heyl, \emph{Physical Review B}, 2015, \textbf{92}, 134204\relax
\mciteBstWouldAddEndPuncttrue
\mciteSetBstMidEndSepPunct{\mcitedefaultmidpunct}
{\mcitedefaultendpunct}{\mcitedefaultseppunct}\relax
\EndOfBibitem
\bibitem[Note5()]{Note5}
In the previous sections, the condition $\xi < 2/\protect \qopname \relax
  o{log}2$ was derived instead. This is simply because there the localized
  material was only on one side of the ergodic grain, whereas now it is on two
  sides.\relax
\mciteBstWouldAddEndPunctfalse
\mciteSetBstMidEndSepPunct{\mcitedefaultmidpunct}
{}{\mcitedefaultseppunct}\relax
\EndOfBibitem
\bibitem[Nandkishore and Gopalakrishnan(2016)]{nandkishore2016general}
R.~Nandkishore and S.~Gopalakrishnan, \emph{arXiv preprint arXiv:1606.08465},
  2016\relax
\mciteBstWouldAddEndPuncttrue
\mciteSetBstMidEndSepPunct{\mcitedefaultmidpunct}
{\mcitedefaultendpunct}{\mcitedefaultseppunct}\relax
\EndOfBibitem
\bibitem[De~Roeck \emph{et~al.}(2016)De~Roeck, Huveneers, M{\"u}ller, and
  Schiulaz]{de2016absence}
W.~De~Roeck, F.~Huveneers, M.~M{\"u}ller and M.~Schiulaz, \emph{Physical Review
  B}, 2016, \textbf{93}, 014203\relax
\mciteBstWouldAddEndPuncttrue
\mciteSetBstMidEndSepPunct{\mcitedefaultmidpunct}
{\mcitedefaultendpunct}{\mcitedefaultseppunct}\relax
\EndOfBibitem
\end{mcitethebibliography}

\appendix
\section{Numerics on the hybridization width}\label{app: numerical evidence for hybridization width}

We return to Section \ref{sec: hybridization function} and we assume the notation used there. We test numerically the (FGR) relationn
$$w \sim \frac{g^2}{\xi}$$
 We always choose the eigenstate $\psi$ of the coupled system to lie in the middle of the spectrum, i.e.\ at  $E = 0$ and consider mainly a random matrix (RM) bath, for which $\xi$ is approximatively the full spectral width. 
The width $w$ can depend a priori on $\rho$, $\xi$, $g$ and $h$, where we recall that $\rho$ is the inverse of the level spacing (at maximal entropy).
For the numerics below we take $h=1$ and $g=0.8$.
As long as $2h$ is significantly large than $g$, we do not expect any crucial dependence on $h$; this turns out to be so indeed (not shown).    
  
\begin{figure}[h!]
\begin{center}
\includegraphics[width=8.5cm,height=5cm]{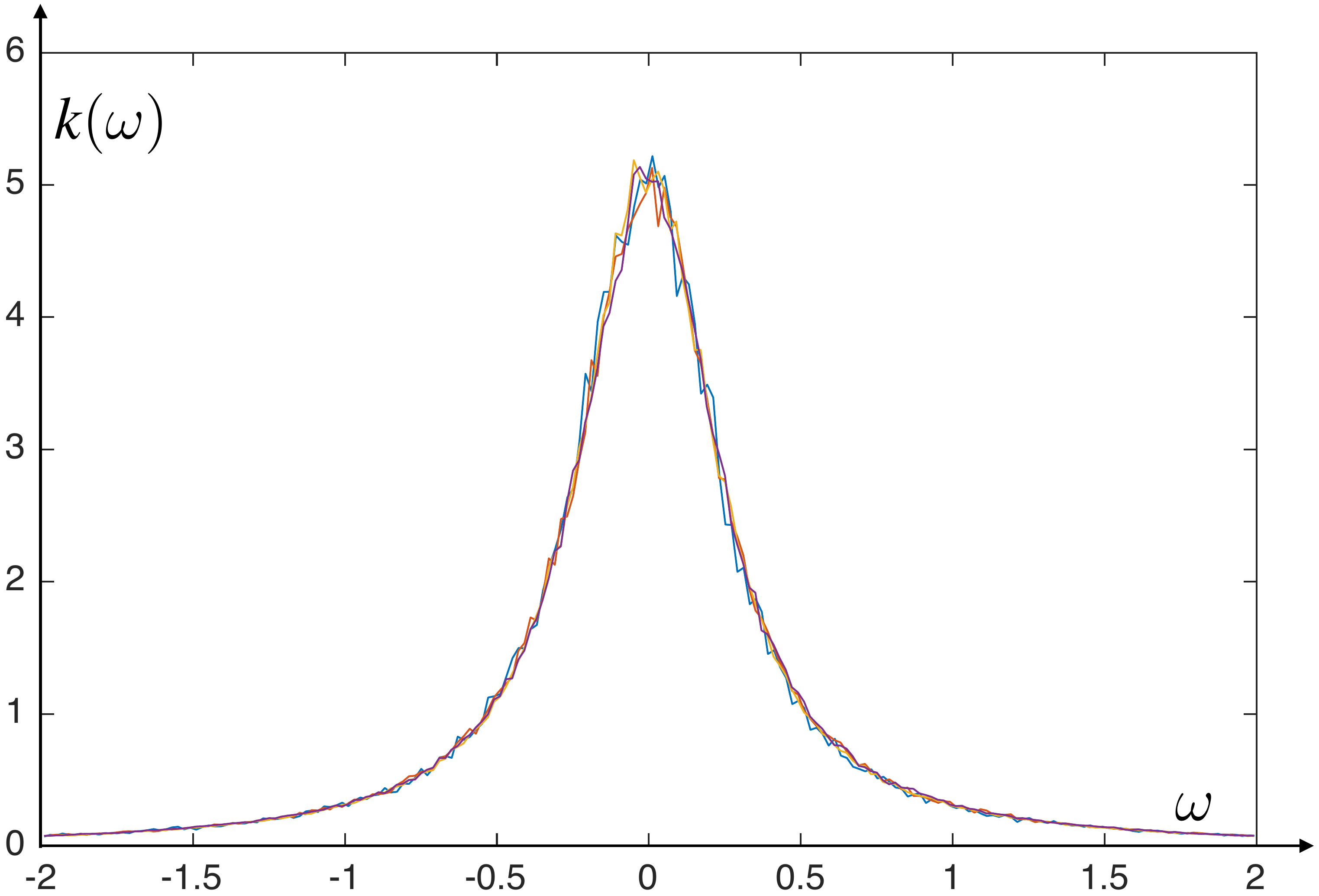}

\bigskip

\includegraphics[width=8.5cm,height=5cm]
{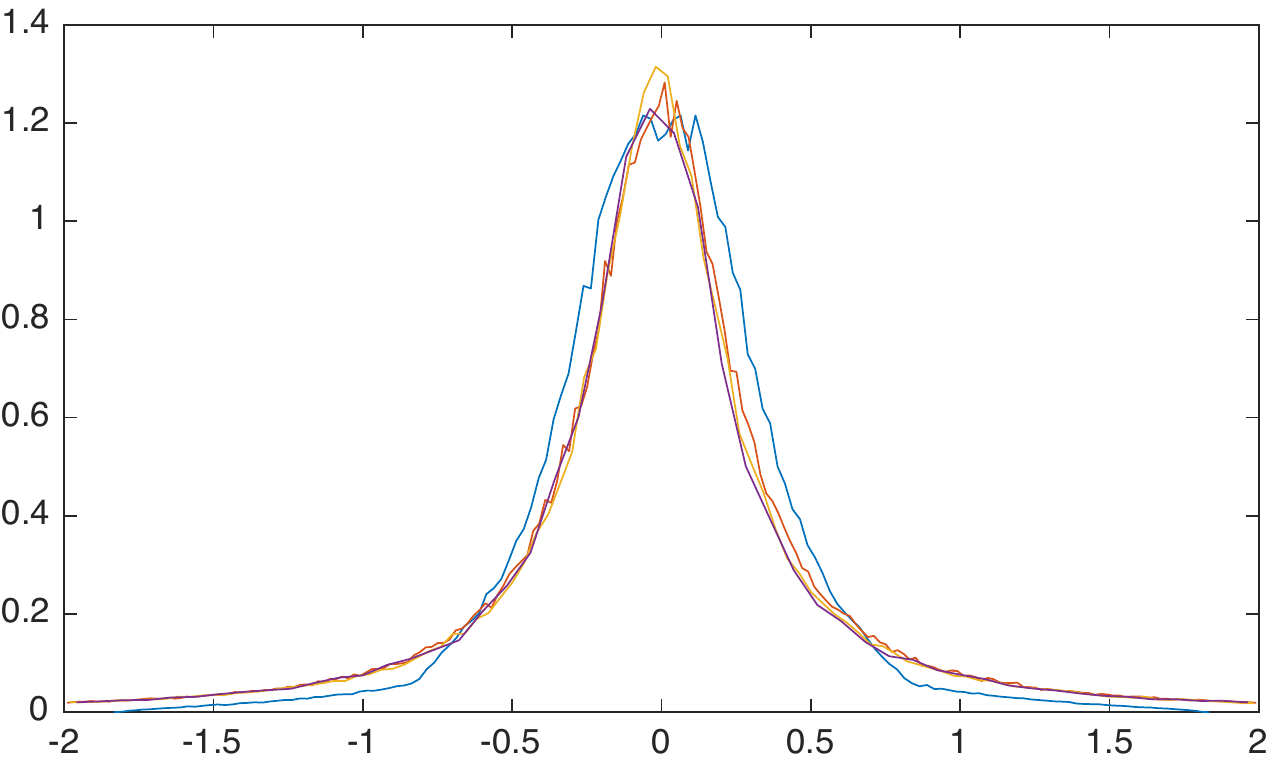}
\end{center}
\caption{
\label{fig: Hybridization level spacing and xi}
Hybridization curves for $4$ different RM baths, by varying the level spacing (upper panel) or the width (lower panel). 
}
\end{figure}

Let us check that $w$ is independent of $\rho$. 
For this, we fix $g$ and $h$, and we consider $4$ different RM baths, say $B_1, \dots B_4$, 
acting on spaces with dimension from $2^{8}$ to $2^{11}$, rescaled in a such a way that $\xi$ remains constant (the width is approximately equal to $\xi = 10$). 
The level spacing is the only variable parameter; it gets divided by $2$ each time we change bath from $B_1$ to $B_4$.
The $4$ corresponding hybridization curves are plotted on the upper panel of figure \ref{fig: Hybridization level spacing and xi}. 
The good matching of the curves indicate that $w$ is indeed independent from $\rho$. 

We next check that $w$ depends on $\xi$ as $1/\xi$. 
For this we repeat the previous numerics, rescaling now the baths so as to keep the level spacing constant (we take it approximately equal to $1/\rho = 0.02$) and varying $\xi$. 
Now the value of $\xi$ gets doubled each time we change bath from $B_i$ to $B_{i+1}$.  
To collapse the curves, we use the scaling relation 
Since a Lorentzian distribution satisfies the scaling 
$\lambda f(\lambda x,\lambda w) = f(x,w)$, the  relation $w \sim 1/\xi$ implies that the curves should collapse by plotting $2^{k-2}f_{B_k} (2^{k-2}x)$ instead of $f_{B_k} (x)$ (the hybridization curves for the bath $B_k$) for $1 \le k \le 4$. 
The result is shown on the lower panel of figure \ref{fig: Hybridization level spacing and xi}. 
We observe again a good matching except for the blue curve corresponding to the smallest bath ($k=1$); 
this discrepancy can be attributed to finite size effects and to the fact that the Lorentzian shape does not need to be strictly verified.  

Thanks to the two previous points, the dependence of $w$ on $g$ as $1/g^2$ now follows from dimensional analysis or simple rescaling of the parameters of the model. 

\begin{figure}[h!]
\begin{center}
\includegraphics[width=8.5cm,height=5cm]{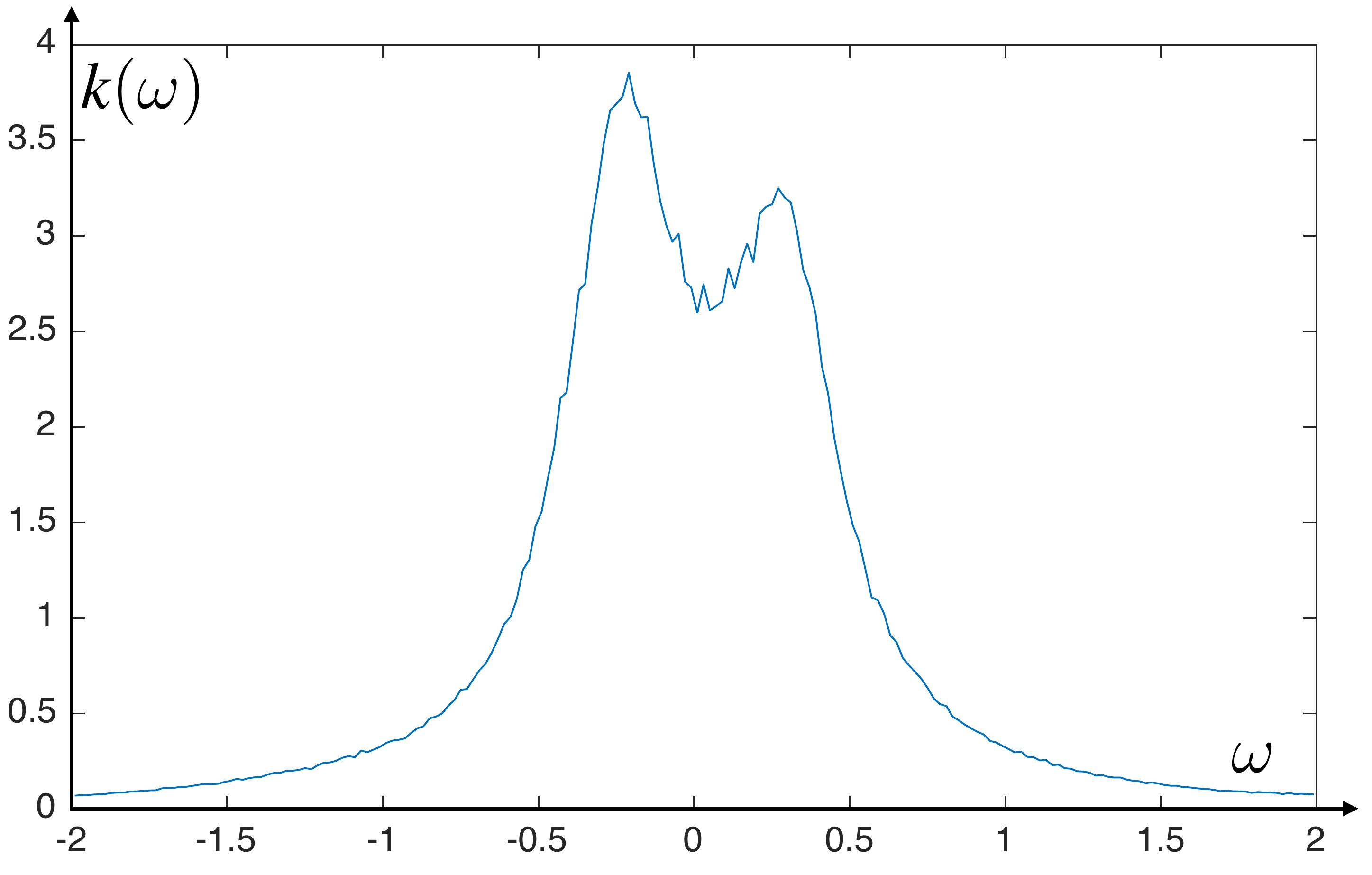}
\end{center}
\caption{
\label{fig: Hybridization Hamiltonian bath}
Hybridization curve for a 11 spins bath with local interactions.}
\end{figure}

Finally, on figure \ref{fig: Hybridization Hamiltonian bath}, we show an example of an hybridization curve for an 11 spins bath with local interactions. 
The Hamiltonian of this bath is the same as the Hamiltonian in eq.~\eqref{Hamiltonian bath}, with  $9$ and $8$ above the summation signs respectively changed to $11$ and $10$. 
We observe a camel-like shape, showing that the Lorentzian distribution should only be taken as an idealization of the hybridization function 
(we also observe some asymmetry in the peaks, which however can be explained by noting that $\mathrm{tr} ({{{H_B}}}^3) \ne 0$). 

\section{Calculations on the Backreaction} \label{app: backreaction}

Here we carry out some calculations that were omitted in Section \ref{sec: correction}. 
First, we evaluate the kernel $K(b,s \str b',s')$ for $b=b'$ in a weak-coupling approximation. 
To that end, we approximate the resolvent by its $g=0$ value $K_0$:
$$
K_0(b,s \str b,s') = g \langle b,s\str  \sigma^x_B  \frac{1}{E'-H_0} \bar P' \sigma^x_s \sigma^x_B  \str b,s' \rangle  
$$
Since $H_0$ does not flip $s$, we can simplify 
$K_0(b,s \str b,s')=\delta_{s,-s'} K_0(b,s \str b,-s)$. Inserting a parition of unity, we write
%
%
\begin{align*}
K_0(b,s \str b,-s) &=   g \sum_{\tilde b}  
\str\langle b\str \sigma^{x} \str \tilde{b}\rangle\str^2
\frac{\chi(\str\Delta\str \geq w)}{\Delta} \\
 &\approx  \int d\omega  \, \chi(\str\Delta(\omega)\str \geq w)
\frac{g v(\omega)}{\Delta(\omega)} 
\end{align*}
where 
$
\Delta  \equiv \omega+2s{{h}}+ \{E(b,-s)-E'\}
$
and we recall that term in $\{\ldots\}$ is restricted to $\str \cdot \str \leq w$. 
In the region where $v$ has its bump, the integrand is typically of size $\tfrac{g}{\xi\max{(h,\xi)}}$, which leads to the estimate $\mathcal{W}\equiv\tfrac{g}{\max{(h,\xi)}}$ for the integral. If the function $v$ were very rough around the cut-off singularity $\Delta(\omega)=0$, then this could change the estimate to $-\mathcal{W}\log\mathcal{W}$, which does not affect the essence of our conclusion.

To get to the contribution to the structure factor \eqref{eq: diagonal contribution v}, we return to the expression
$$
\langle P\psi \str \sigma^x_B \str \bar P'\psi' \rangle =\sum_{b,b',s,s'} \langle P\psi \str b,s \rangle  K(b,s \str b',s')   \langle b',s' \str  P'\psi' \rangle  
$$
Restricting to $b=b'$, replacing $K$ by $K_0$ and $ \langle P\psi \str b,s \rangle, \langle b',s' \str  P'\psi' \rangle $ by the random expressions from \eqref{eq: refined proposal}, we get the result  by a central-limit calculation, as in Section \ref{sec: new structure factor}.

As mentioned in Section \ref{sec: correction}, similar considerations also lead to a derivation of the FGR, which is phrased here as the fact that $\norm \bar P \psi \norm \sim 1$, i.e.\ we are truncating the eigenstates exactly around their bump. Approximating the resolvent again by its lowest-order expression;
$$
\bar P \psi \approx g \sum_{b,s}  \frac{\sqrt{k_0(\omega)} \eta(b,s)}{\sqrt\rho} \frac{1}{E-H_0} \bar P \sigma^x_S \sigma^x_B  \str b,s \rangle  
$$
with $\omega=E-E(b,s)$, we can calculate (the expectation value of) $\norm \bar P \psi \norm^2$ as
$$
\norm \bar P \psi \norm^2 \approx  g^2\sum_{b,s,b',s'}   \frac{{k_0(\omega)}\chi(\str\omega'\str\geq w) }{(\omega')^2 \rho }   \str\langle b',s'\str \sigma^x_B  \str b,s \rangle \str^2    
$$
where $\omega'=E-E(b',s')$. Plugging the ETH expression for the matrix elements of $\sigma^x_B$, we get 
$$
\norm \bar P \psi \norm^2 \approx  
g^2 \int \d\omega \d \omega'   \frac{k_0(\omega)\chi(\str\omega'\str\geq w) }{(\omega')^2 }  v(\omega'-\omega)
$$
We use that $\int k_0 \sim 1$ (with width $w$) and we approximate $v$ as a bump with width $\xi \gg w$ (FGR condition). Then the above integral yields indeed $g^2/\xi$. Most importantly, this conclusion is not affected if $v$ had additional narrow peaks with small weight, cf.\ the splitting $v=v_{\text{sm}}+v_{\text{irr}}$ obtained in Section \ref{sec: correction}.

\end{document}